\documentclass[twocolumn,twocolappendix]{aastex63}

\usepackage{listings}
\usepackage{amsmath}

\graphicspath{{./}{figures/}}
\shorttitle{CLIMBER}
\shortauthors{Pearl, et al.}

\newcommand{\lcdm}{$\mathrm{\Lambda}$CDM}
\newcommand{\ssfruv}{$\mathrm{sSFR_{UV}}$}

\begin{document}

\title{CLIMBER: Galaxy-Halo Connection Constraints from Next-Generation Surveys}

\author[0000-0001-9820-9619]{Alan N.\ Pearl}
\affiliation{Department of Physics and Astronomy, University of Pittsburgh, Pittsburgh, PA 15260}
\author[0000-0001-5063-8254]{Rachel Bezanson}
\affiliation{Department of Physics and Astronomy, University of Pittsburgh, Pittsburgh, PA 15260}
\author[0000-0002-6443-7186]{Andrew R.\ Zentner}
\affiliation{Department of Physics and Astronomy, University of Pittsburgh, Pittsburgh, PA 15260}
\author[0000-0001-8684-2222]{Jeffrey A.\ Newman}
\affiliation{Department of Physics and Astronomy, University of Pittsburgh, Pittsburgh, PA 15260}
\author[0000-0003-4700-663X]{Andy D.\ Goulding}
\affiliation{Department of Astrophysical Sciences, Princeton University, Princeton, NJ 08544}
\author[0000-0001-7160-3632]{Katherine E.\ Whitaker}
\affiliation{Department of Astronomy, University of Massachusetts, Amherst, MA 01003}
\author[0000-0001-9487-8583]{Sean D.\ Johnson}
\affiliation{Department of Astronomy, University of Michigan, Ann Arbor, MI 48109}
\author[0000-0002-5612-3427]{Jenny E.\ Greene}
\affiliation{Department of Astrophysical Sciences, Princeton University, Princeton, NJ 08544}

\begin{abstract}
    In the coming decade, a new generation of massively multiplexed spectroscopic surveys, such as PFS, WAVES, and MOONS, will probe galaxies in the distant universe in vastly greater numbers than was previously possible. In this work, we generate mock catalogs for each of these three planned surveys to help quantify and optimize their scientific output. To assign photometry into the UniverseMachine empirical model, we develop the Calibrating Light: Illuminating Mocks By Empirical Relations (CLIMBER) procedure using UltraVISTA photometry. Using the published empirical selection functions for each aforementioned survey, we quantify the mass completeness of each survey. We compare different targeting strategies by varying the area and {targeting} completeness, and quantify how these survey parameters affect the uncertainty of the two-point correlation function. We demonstrate that the PFS and MOONS measurements will be primarily dominated by cosmic variance, not shot noise, motivating the need for increasingly large survey areas. On the other hand, the WAVES survey, which covers a much larger area, will strike a good balance between cosmic variance and shot noise. {For a fixed number of targets, a 5\% increased survey area (and $\sim$5\% decreased completeness) would decrease the uncertainty of the correlation function at intermediate scales by 0.15\%, 1.2\%, and 1.1\% for our WAVES, PFS, and MOONS samples, respectively. Meanwhile, for a fixed survey area, 5\% increased targeting completeness improves the same constraints by 0.7\%, 0.25\%, and 0.1\%.} All of the utilities used to construct our mock catalogs and many of the catalogs themselves are publicly available.
\end{abstract}

\section{Introduction}

The \lcdm{} model of cosmology has been widely accepted for decades, and its parameters are now known to quite high precision \citep{Planck:2018}. Within this framework, it is assumed that galaxies form inside dark matter halos \citep{White:Rees:1978, Blumenthal:1984}.
While halos can be accurately modeled through gravity-only simulations \citep[e.g., Bolshoi-Planck;][]{Klypin:2016}, the fine details of galaxy formation are strongly influenced by baryonic physics, which poses a serious challenge for theoretical models of galaxy evolution to tackle.

In recent years, several ongoing projects have made great advancements to including baryonic physics in hydrodynamic simulations of galaxies in a cosmological context. These projects include but are not limited to IllustrisTNG \citep{Nelson:2019}, Evolution and Assembly of GaLaxies and their Environments \citep[EAGLE;][]{Schaye:2015, Crain:2015}, Feedback In Realistic Environments \citep[FIRE;][]{Hopkins:2018}, and Simba \citep{Dave:2019}. However, it is still computationally prohibitive to resolve the small scales needed to simulate the processes that regulate star formation. Therefore, all of these hydrodynamic simulations still include analytic approximations for these small-scale processes. It is possible to approximate the rates of processes like gas cooling and star formation using semi-analytic models \citep[SAMs; e.g.,][]{White:Frenk:1991, Somerville:2015} which trace dark matter halos through gravity-only simulations and map baryonic physics into these halos using analytic scaling relations. SAMs have contributed significantly to our knowledge of galaxy formation, despite challenges disentangling various physical processes that produce degenerate observations.

Alternatively, many studies of galaxy evolution and cosmology use empirical models to populate galaxies on top of dark matter halos (i.e., the galaxy-halo connection; see \citealt{Wechsler:Tinker:2018} for an extensive review). Methods such as the halo occupation distribution \citep[HOD; e.g.,][]{Berlind:Weinberg:2002, Zheng:2007, Hearin:2016} 
or abundance-matching \citep[e.g.,][]{Kravtsov:2004, Hearin:Watson:2013} 
are most commonly used to statistically match the stellar masses of galaxies to the masses of their host halos. Since the number density and clustering of halos are strongly dependent on halo mass \citep{Press:Schechter:1974, Kaiser:1984, Bond:1991, Mo:White:1996, Zentner:2007}, 
these models are informed through observations of the stellar mass function and the two-point correlation function \citep{Zehavi:2005, Reddick:2013, Wang:2019}.

Through analytic empirical models, one can infer the stellar-to-halo mass relation (SHMR). The SHMR has provided us valuable insight into the masses of halos that are most efficient at forming stellar mass. At its peak, stellar mass can account for up to roughly 5\% of the total mass of Milky Way-mass halos, which is approximately 30\% of the cosmic baryon mass fraction. However, this star formation efficiency drops dramatically at both lower and higher halo masses. This is caused by various processes causing star formation to shut off (i.e., quench) through heating or removing the gas that was fueling the star formation. Low-mass quenching is often attributed to stellar feedback \citep{Fierlinger:2016} and satellite stripping \citep{Guo:2011}, while high-mass quenching is primarily attributed to active galactic nucleus (AGN) feedback \citep{Fabian:2012}. However, the dependence of these processes on secondary halo properties and redshift is poorly constrained by existing datasets.

Most of our constraints on the galaxy-halo connection come from low-redshift surveys such as the Sloan Digital Sky Survey \citep[SDSS;][]{Blanton:2017}. While it is commonly assumed that the SHMR does not evolve strongly with redshift, it is particularly difficult to probe the same range of halos at high redshifts because these surveys quickly lose faint, low-mass galaxies and massive galaxies are rare. Extending our empirical constraints on the galaxy-halo connection to the high-redshift universe, where star formation rates (SFRs) were higher and galaxy populations were rapidly evolving, should have profound implications on our knowledge of galaxy evolution.

With the advent of highly multiplexed spectrographs being used on large telescopes, thousands of spectra will be simultaneously measured, a number of spectroscopic surveys will begin to map the distant universe to an unparalleled degree over the next decade. These surveys will probe the evolution of the precise statistical distribution of galaxies at earlier cosmological times than previously possible. Interpreting these types of datasets, however, is particularly challenging due to the systematic sampling that is more easily avoidable in the nearby universe. Utilizing this new information to place constraints on the galaxy-halo connection will require careful planning of survey designs and new theoretical frameworks.

In this paper, we present a procedure for mapping photometric properties (flux and colors) onto physical properties from the UniverseMachine empirical model \citep{Behroozi:2019}. We refer to this procedure as Calibrating Light: Illuminating Mocks By Empirical Relations (CLIMBER). We use this procedure to construct mock galaxy catalogs, which we use to investigate the mass completeness and statistical constraints that will be available from several future massively multiplexed spectroscopic galaxy surveys: the Prime Focus Spectrograph Galaxy Evolution Survey \citep[PFS;][]{Takada:2014}, the Guaranteed Time Observation Extragalactic Survey of the Multi-Object Optical and Near-infrared Spectrograph for the Very Large Telescope \citep[MOONS;][]{Maiolino:2020}, and the Wide Area Vista Extragalactic Survey-Deep \citep[WAVES;][]{Driver:2016}. For each survey, we quantify its performance and make recommendations about future extensions to improve its constraining power on the galaxy-halo connection.

This paper is organized as follows: Section~\ref{sec:mocksurvey} explains the procedure we followed to construct our mock galaxy catalog from the fiducial UniverseMachine model (with more details in Appendix~\ref{sec:climber-details}) and discusses selection functions that we place to construct mock surveys of various galaxy populations. In Section~\ref{sec:hod-formulation}, we formulate our conservative HOD model. In Section~\ref{sec:constraints}, we present mock measurements of number density and the two-point correlation function, which are the primary constraints of this model. In Section~\ref{sec:analysis}, we present projected constraints of the two-point correlation function and HOD models through Markov chain Monte Carlo (MCMC) fits, for a variety of survey parameters. We give a brief discussion of our conclusions in Section~\ref{sec:conclusions}.

The cosmological assumptions used in each step of generating our mock catalog were made self-consistently. Bolshoi-Planck, the UniverseMachine, and all of our following calculations use a Planck-tuned \lcdm{} cosmology with parameters given in Table~\ref{tab:cosmo}. Although the stellar masses and SFRs from UltraVISTA \citep{Muzzin:2013} assumed a slightly different cosmology, their dependence on $h$ has been corrected to match our assumption. Note that all halo masses refer to the virial mass of the halo, and we do not use $h$-scaled units with the exception of $h^{-1}$~Mpc for distance.

\startlongtable
\begin{deluxetable}{c|c|c}
\tablecaption{Cosmological parameters \label{tab:cosmo}}
\tablehead{
\colhead{Parameter} & \colhead{Value} & \colhead{Description}
}
\startdata
$h$ & 0.678 & Hubble parameter \\
$\Omega_\Lambda$ & 0.693 & density parameter for dark energy
\\
$\Omega_m$ & 0.307 & density parameter for total matter \\
$\Omega_b$ & 0.048 & density parameter for baryonic matter \\
$n_s$ & 0.96 & normalization of the Power spectrum \\
$\sigma_8$ & 0.823 & amplitude of mass density fluctuation \\
\enddata
\end{deluxetable}

\section{Building the Empirically Calibrated Mock Surveys}
\label{sec:mocksurvey}

\begin{figure*}[ht!]
\includegraphics[width=\textwidth]{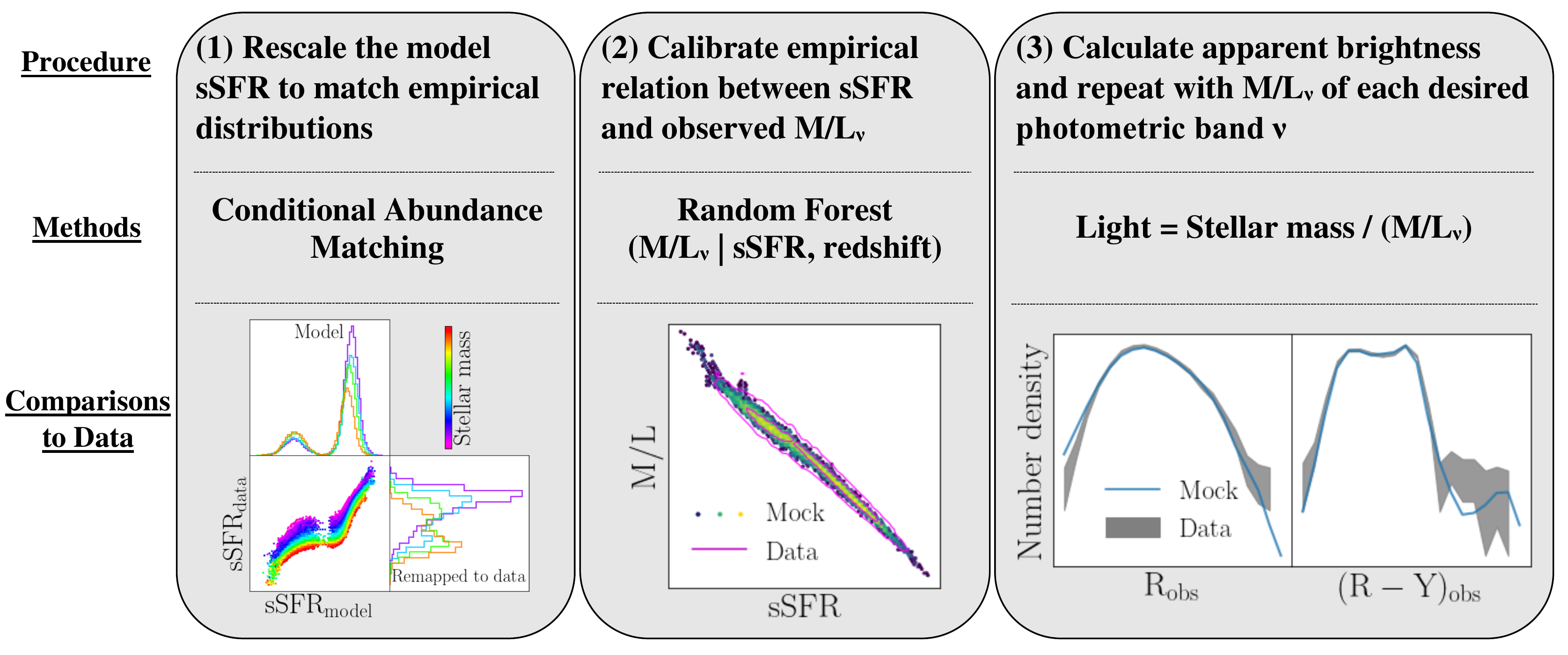}
\caption{Visualization of our calibration procedure (CLIMBER) developed to assign the brightness and color of each mock galaxy taken from the UniverseMachine empirical model. Note that we abbreviate specific SFR (sSFR; i.e., SFR divided by stellar mass) and stellar mass-to-light ratio ($\rm M/L_\nu$), where $\rm \nu$ represents the effective observed-frame frequency of a photometric band.
\label{fig:flowchart}}
\end{figure*}

To construct a realistic mock galaxy catalog, we start from the UniverseMachine \citep{Behroozi:2019} empirical model, which is calibrated to reliably reproduce a very large number of statistics of galaxy populations from $0<z<10$. However, this model lacks a crucial element needed to test empirical selection functions: the apparent brightness of each galaxy in the observed-frame wavelengths of photometric filters. We, therefore, calibrate these model galaxies to photometry from the UltraVISTA survey \citep{Muzzin:2013} using a combination of abundance matching and random forest mapping to calculate observed mass-to-light ratios (see Section~\ref{sec:climber}).

\subsection{UniverseMachine}
\label{sec:um}

The UniverseMachine \citep{Behroozi:2019} is a sophisticated empirical galaxy-halo connection model of 44 parameters, which were iteratively fit to 1069 observed data points across a redshift range of $0<z<10$. It traces each dark matter halo in a gravity-only simulation and assigns a SFR to the galaxy at the center of the halo, assuming that star formation correlates with dark matter assembly. The model thereby tracks the accumulation of stellar mass of each galaxy over its entire formation history. The SFR of each galaxy is drawn from an empirically motivated distribution, which is the sum of two log-normal distributions, representing a quenched and star-forming population. The quenched fraction, as well as the center and width of the star-forming distribution, are parameterized by analytic expressions dependent on halo mass and redshift. In order to impose some assembly correlation, the SFR is not randomly drawn from the model distribution, but instead weighted such that higher SFRs are more likely to be assigned to halos with greater mass accretion rates. For an excellent visual summary of this procedure, see Figure~1 of \citet{Behroozi:2019}.

The UniverseMachine DR1 derives its halo catalog from the Bolshoi-Planck cosmological N-body simulation \citep{Klypin:2016}. The UniverseMachine then provides mock galaxy properties such SFR and stellar mass (note that this is the ``live'' stellar mass, which is the integral of the star formation history subtracted by the mass returned to the interstellar medium) into each snapshot of this simulation. By piecing together these snapshots, the UniverseMachine has been tuned to reproduce many observables, such as stellar mass functions, two-point correlation functions, the star-forming main sequence, quenched fractions, environmental quenching, and more.

Before using the UniverseMachine to generate mock surveys, one needs to define the empirical properties of each mock galaxy to impose selection functions.
This has previously been done by performing stellar population synthesis over each star formation history to fit the UniverseMachine to UV luminosity functions and UVJ quenching classifications. However, the UniverseMachine is only tuned to reproduce global star formation histories. Because our goal is to apply targeting strategies, the distribution of colors and fluxes must be reliable, and we, therefore, empirically calibrate a mapping from stellar mass and SFR (the \verb|obs_sm| and \verb|obs_sfr| columns) to the brightness in various photometric filters, as explained in Section~\ref{sec:climber}.

\subsection{CLIMBER}
\label{sec:climber}

Most spectroscopic galaxy surveys have well-defined selection functions based on previously taken photometric data. Therefore, we need a method of predicting the observed light of UniverseMachine galaxies at multiple wavelengths to understand the representation of properties of the targeted galaxies (e.g., mass completeness) of these surveys. For this reason, we have developed a procedure to assign mock apparent magnitudes informed by an observational dataset. We refer to this procedure as Calibrating Light: Illuminating Mocks By Empirical Relations (CLIMBER).

In CLIMBER, we utilize the tight correlation between the mass-to-light ratio and color of a galaxy \citep[e.g.,][]{Bell:deJong:2001}. Analogously, we map sSFR to the mass-to-light ratio for each mock galaxy via random forest regression. This can be trained by any observational dataset with the desired mass, redshift, and photometric coverage. The data shown in this paper have been calibrated to UltraVISTA photometry \citep{Muzzin:2013}, but CLIMBER is a generalizable, flexible procedure that can be applied to any dataset that includes mappings between empirical fluxes and physical galaxy properties.

While the scaling relation between total SFR and stellar mass of star-forming galaxies (the star-forming main sequence) has been measured out to $z \sim 5$ \citep{Kajisawa:2010, Whitaker:2014, Tomczak:2016, Leslie:2020}, the evolution of SFRs for galaxies that fall off this relation is poorly constrained due to the difficulty of measuring low SFRs \citep{Leja:2019}. For this reason, the UniverseMachine only evolves the star-forming SFR distribution with redshift. In contrast, the specific SFR (sSFR) of each quiescent galaxy in the UniverseMachine was simply drawn from a non-evolving log-normal distribution of $10^{-11.8}$~yr$^{-1} \pm 0.36$~dex. While this empirically matches the local universe, it likely underestimates SFRs of high-redshift quiescent populations. This assumption does not greatly influence the accumulation of stellar mass modeled in the UniverseMachine, but it presents a problem for assigning the luminosities of quiescent galaxies. We solve this by rescaling the sSFR distributions via conditional abundance matching, as discussed in greater detail in Appendix~\ref{sec:climber-details}.

The most crucial decision one needs to make before running CLIMBER is in choosing an sSFR proxy from the calibration dataset. This proxy must (1) approximately conserve rank-ordering with true sSFR (at fixed stellar mass), (2) produce a tight, negative correlation with mass-to-light ratio, and (3) have a high detection fraction in the full galaxy population -- quenched and star-forming galaxies alike. From UltraVISTA, we chose to use the specific ultraviolet SFR (\ssfruv{}), which is the SFR inferred from ultraviolet bands, divided by stellar mass derived from SED fitting. These SEDs were fit by Fitting and Assessment of Synthetic Templates \citep[FAST;][]{Kriek:2009} using the \citet{Chabrier:2003} initial mass function, an exponentially declining star formation history, and the \citet{Bruzual:Charlot:2003} stellar population synthesis model. Other sSFR proxies may be useful in different datasets. For example, we considered using sSFRs directly from SED fits, but the grid-based values fit by FAST were sampled too sparsely and therefore provide a poor mapping between physical properties and flux. 

See Figure~\ref{fig:flowchart} for a flow chart visualization of the CLIMBER procedure. To summarize, we first perform conditional abundance matching from the model sSFR to match the empirical sSFR distribution. Then, we train the mapping from sSFR to an observed mass-to-light ratio using random forest. Finally, we convert the UniverseMachine stellar mass values to luminosities via the predicted mass-to-light ratios. For further details and analysis of our procedure, see Appendix~\ref{sec:climber-details}.

\subsection{Mock Survey Selections}
\label{sec:selections}

The product of CLIMBER is a mock realization of the universe in the form of a light cone, which can be iterated over random origins and orientations in the Bolshoi-Planck cube. We conduct mock surveys over these light cones by performing cuts that imitate the selection functions of several next-generation surveys. We then analyze many realizations of each mock survey to try to determine the uncertainty of the number density and two-point correlation function (see Sections~\ref{sec:smf} and~\ref{sec:2pcf}) that will be measured.

In this work, we ignore the intricate details of survey geometries, overlapping pointings, and fiber collisions, which would require targeting strategies that are not yet finalized (however, our mock catalogs will be an extremely useful tool for running targeting simulations and analyzing the systematics they produce). We define our survey geometry by a square in angular coordinates and remove a random subsample to account for {targeting completness}, primarily due to fiber collisions. Our two survey parameters are thus sky area and {targeting} completeness.

We implement this selection using $|\alpha| < \alpha_{\rm max}$ and $|\delta| < \delta_{\rm max}$ where $\alpha_{\rm max} = \delta_{\rm max}$. For small angles, the solid angle area is approximately $\Omega \approx 4 \alpha_{\rm max} \delta_{\rm max}$ (in radians/steradians), but to be precise, we calculate $\alpha_{\rm max}$ and $\delta_{\rm max}$ by inverting Equation~\ref{eq:angle-to-area}.
\begin{equation} \label{eq:angle-to-area}
\begin{split}
    \Omega 
    & = \int_{-\alpha_{\rm max}}^{\alpha_{\rm max}} d\alpha \int_{-\delta_{\rm max}}^{\delta_{\rm max}} d\delta \cos(\delta) \\[8pt]
    & = 4(\alpha_{\rm max}) \sin(\delta_{\rm max})
\end{split}
\end{equation}

\startlongtable
\begin{deluxetable*}{c|c|c|c|c|c}
\tablecaption{Survey parameters \label{tab:survey-params}}
\tablehead{
\colhead{Name} & \colhead{Area (sq.\ deg)} & \colhead{Completeness} & \colhead{Redshift} & \colhead{Magnitude limits} & \colhead{References}
}
\startdata
WAVES & 66 & 95\% & $0.2 < z < 0.8$ & $m_z < 21.25$ & \citet{Driver:2016}\\[3pt]
PFS ($z < 1$) & 12 & 70\% & $0.7 < z < 1.0$ & $m_Y < 22.5$ \& $m_J < 22.8$ & \citet{Takada:2014}\\
PFS ($z > 1$) & 12 & 70\% & $1.0 < z < 1.7$ & $m_J < 22.8$ & \citet{Takada:2014}\\[3pt]
MOONS\tablenotemark{a} ($z \sim 1$) & 4 & 72.5\% & $0.9 < z < 1.1$ & $m_H < 23.0$ & \citet{Maiolino:2020}\\
MOONS\tablenotemark{a} ($z \sim 1.5$) & 4 & 72.5\% & $1.2 < z < 1.7$ & $m_H < 23.5$ & \citet{Maiolino:2020}\\
MOONS\tablenotemark{a} ($z \sim 2$) & 4 & 72.5\% & $2.0 < z < 2.6$ & $m_H < 24.0$ & \citet{Maiolino:2020}\\
\enddata
\tablenotetext{a}{These parameters assume that MOONS implements the Xswitch strategy. If the more efficient Stare strategy is used in the VIDEO fields, the total area would increase to 7 sq.\ deg and the average {targeting} completeness would drop to 71.4\%.
}
\end{deluxetable*}

\begin{figure}[ht!]
\includegraphics[width=0.47\textwidth]{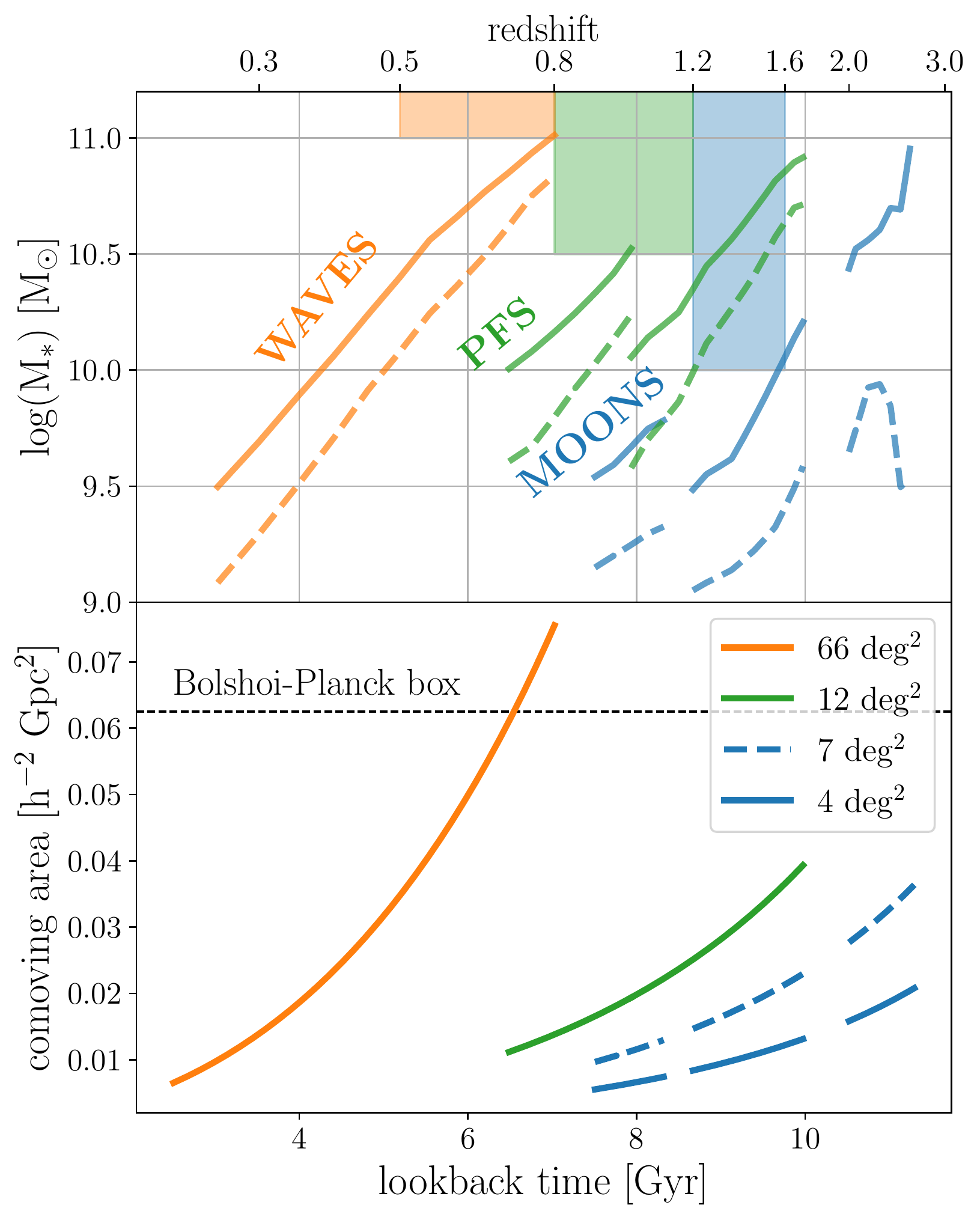}
\caption{Mass completeness (upper panel) and field size (lower panel) for the targeting strategies (given in Table~\ref{tab:survey-params}) of PFS, WAVES, and MOONS as a function of redshift. In the upper panel, we include 99\% (solid lines) and 90\% (dashed lines) completeness limits. These values are averaged over 25 mock catalog realizations. Color-coded bands indicate the mass-complete samples used in this analysis (see Table~\ref{tab:galaxy-samples}). In the lower panel, we plot the comoving area of each field, with sky areas taken from Table~\ref{tab:survey-params}.
\label{fig:mass-completeness}}
\end{figure}

To perform any type of scientific study on a sample of galaxies, its selection function must be well understood in terms of physical properties. By imposing the published magnitude limits (see Table~\ref{tab:survey-params}) of the WAVES, PFS, and MOONS surveys on galaxies in our mock, we can test the fraction of galaxies at a given mass that is included in the selection function to test how well-represented they will be in the survey. For each survey, we show the 90\% and 99\% mass-completeness limit as a function of redshift in Figure~\ref{fig:mass-completeness}. The color-coded bands in this figure enclose the three galaxy populations that we further analyze in this paper by calculating mock observables (Section~\ref{sec:constraints}) to constrain our HOD model (Section~\ref{sec:hod-formulation}). The mass thresholds and effective redshifts of these samples are listed in Table~\ref{tab:galaxy-samples}. These cuts are almost entirely above the respective 99\% mass-completeness limits, which means these surveys should observe representative samples.

Figure~\ref{fig:mass-completeness} additionally shows the comoving area probed by each survey as a function of redshift, in comparison to that of the Bolshoi-Planck simulation, which is a periodic cube of side length $250~h^{-1}~{\rm Mpc}$. Note that WAVES reaches a slightly larger comoving area at the high-redshift end, which may cause the cosmic variance in our mocks to be slightly underestimated, due to the high probability of resampling the same galaxies across realizations. While this should not affect our primary conclusions, a more precise analysis of the cosmic variance in WAVES should use a larger simulation than Bolshoi-Planck.

\startlongtable
\begin{deluxetable*}{c|c|c|c|c|c}
\tablecaption{Galaxy samples \label{tab:galaxy-samples}}
\tablehead{
\colhead{Name} & \colhead{Mass threshold ($M_\odot$)} & \colhead{Mass completeness} & \colhead{Redshift range} & \colhead{Effective redshift} & \colhead{Mean sample size}
}
\startdata
WAVES & $10^{11}$ & 99.903\% & $0.5 < z < 0.8$ & 0.647 & 33,{}583\\
PFS & $10^{10.5}$ & 99.997\% & $0.8 < z < 1.2$ & 0.979 & 61,{}307\\
MOONS & $10^{10}$ & 99.744\% & $1.2 < z < 1.6$ & 1.367 & 41,{}661\\
\enddata
\end{deluxetable*}

\section{HOD Formulation}
\label{sec:hod-formulation}

\begin{figure}[ht!]
\includegraphics[width=0.47\textwidth]{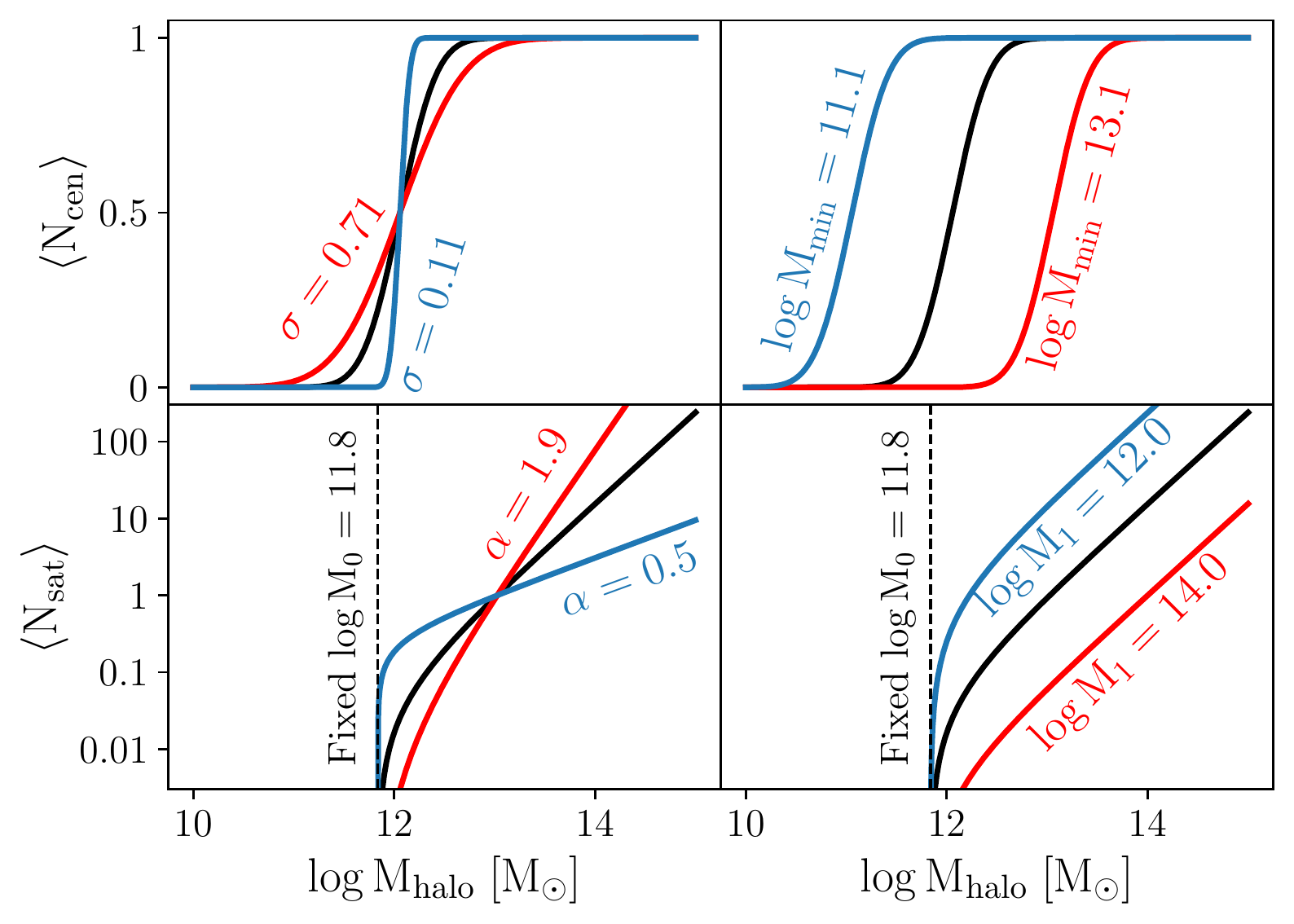}
\caption{Mean occupation functions of centrals (Equation~\ref{eq:hodcen}; top panels) and satellites (Equation~\ref{eq:hodsat}; bottom panels) per halo in our HOD model. The total number of galaxies per halo is the sum of $N_{\rm cen}$ and $N_{\rm sat}$. Black curves are plotted with our fiducial set of parameters for our PFS galaxy sample. We demonstrate the effect of each model parameter by varying them one at a time, as labeled. Note that our conservative HOD model (Section~\ref{sec:conservative-hod}) always maintains a constant $M_{\rm min} / M_1$ ratio, and conserves total number density by automatically updating $\log M_{\rm min}$ and $\log M_1$ to account for any change in $\sigma$ or $\alpha$.
\label{fig:hod}}
\end{figure}

\subsection{The HOD}
\label{sec:hod}

In this paper, we will predict the level of constraints that several upcoming surveys will place on the galaxy-halo connection. To quantify these constraints, we use the halo occupation distribution (HOD), which has been a standard way to measure the galaxy-halo connection in magnitude limited surveys for nearly two decades \citep{Berlind:Weinberg:2002}.

The HOD prescribes the mean number of galaxies above a mass or luminosity threshold per halo. This formalism is very popular due to its simplicity and utility for galaxy clustering predictions. We use the HOD parameter convention introduced by \citet{Zheng:2007}. Under this formalism, we describe the expected number of central and satellite galaxies per halo above a stellar mass threshold, $M_{\ast{\rm thresh}}$, as 
\begin{equation} \label{eq:hodcen}
    \langle N_{{\rm cen}} \rangle = \frac{1}{2}\left(1 + {\rm erf}{\left(\frac{\log\left(M_h / M_{\rm min}\right)}{\sigma}\right)}\right)
\end{equation}
and
\begin{equation} \label{eq:hodsat}
    \langle N_{{\rm sat}} \rangle = \left(\frac{M_h - M_0}{M_1}\right)^\alpha,
\end{equation}
where we do not assume any functional forms for the redshift and $M_{\ast{\rm thresh}}$ dependence of the free parameters $M_{\rm min}$, $\sigma$, $M_0$, $M_1$, and $\alpha$. Instead, we fit the HOD independently to each galaxy population of interest.

We plot these equations in Figure~\ref{fig:hod} using the fiducial parameters for our PFS sample and demonstrate how varying these parameters varies the number of galaxies per halo.

The parameters controlling $\langle N_{\rm cen} \rangle$ are the characteristic halo mass $M_{\rm min}$ and the characteristic spread $\sigma$.
The parameters controlling $\langle N_{\rm sat} \rangle$ are the minimum halo mass $M_0$, 
the characteristic halo mass $M_1$, and the power-law slope $\alpha$. 

While the mean occupation function is a deterministic function of the HOD parameters, note that the number of galaxies assigned to each halo is stochastically drawn from a distribution around this mean. We draw from a Bernoulli distribution (1 or 0) for central galaxies and a Poisson distribution for satellites. Therefore, this induces some stochasticity in the number density and correlation function when populating a simulation of finite volume according to our HOD.

The HOD is constrained by quantities that probe the mass of the underlying halo population: abundance and clustering. To quantify clustering, we measure the two-point correlation function, $w_{\rm p}(r_{\rm p})$ (see Section~\ref{sec:2pcf}). We quantify abundance with the number density of galaxies above the stellar mass threshold, $n$, which is related to the SMF (see Section~\ref{sec:smf}) by
\begin{equation} \label{eq:smf}
    n = \int_{M_{\ast{\rm thresh}}}^\infty \Phi(M_\ast) dM_\ast.
\end{equation}

The HOD can be calculated directly from the UniverseMachine by counting the average number of galaxies above the threshold in each halo. We fit Equations~\ref{eq:hodcen} and~\ref{eq:hodsat} to this calculation in narrow $M_h$ bins to obtain fiducial HOD parameters for our WAVES, PFS, and MOONS galaxy samples. We list each fiducial HOD parameter, as well as the average number density $n$ and satellite fraction $f_{\rm sat}$ in Table~\ref{tab:hodparams}.

\startlongtable
\begin{deluxetable*}{c|c|c|c|c|c|c|c|c}
\tablecaption{Fiducial HOD parameters (UniverseMachine ``truths'') \label{tab:hodparams}}
\tablehead{
\colhead{Sample} & \colhead{$n$ ($h^{3}~$Mpc$^{-3}$)} & \colhead{$M_1/M_{\rm min}$} & \colhead{$\sigma$} & \colhead{$\alpha$} & \colhead{$\log M_{\rm min}$} & \colhead{$\log M_1$} & \colhead{$\log M_0$} & \colhead{$f_{\rm sat}$}
}
\startdata
(\it Parameter type) & (\it Fixed) & (\it Fixed) & (\it Free) & (\it Free) & (\it Derived) & (\it Derived) & (\it Fixed) & (\it Derived)\\
WAVES & 9.992$\times10^{-4}$ & 5.229 & 0.736 & 1.291 & 12.956 & 13.674 & 12.176 & 0.190\\
PFS & 4.838$\times10^{-3}$ & 8.691 & 0.407 & 1.188 & 12.058 & 12.997 & 11.838 & 0.221\\
MOONS & 7.619$\times10^{-3}$ & 9.132 & 0.220 & 1.196 & 11.740 & 12.701 & 11.655 & 0.222\\
\enddata
\end{deluxetable*}

\subsection{The Conservative HOD Model}
\label{sec:conservative-hod}

Combining information from multiple sources, we expect very strong empirical constraints on the number density of most galaxy populations. Even in our mock surveys alone, we measure $n$ to the precision of 1 to 4\%, which is an order of magnitude smaller than the fractional error of $w_{\rm p}$ at large scales. Therefore, if allowed to freely vary, the number density causes a near-degeneracy between the HOD parameters it is sensitive to, increasing the difficulty of calculating constraints through MCMC with little gain.
We, therefore, set the fiducial value of $n$ as a hard prior and only consider the HOD parameter-space that conserves the number density from the UniverseMachine.
To do this, we integrate Equations~\ref{eq:hodcen} and~\ref{eq:hodsat} with the halo mass function $\Phi(M_h)$ to obtain
\begin{equation} \label{eq:numcen}
\begin{split}
n_{{\rm cen}} = \int_0^\infty \langle N_{{\rm cen}} \rangle \Phi(M_h) dM_h
\end{split}
\end{equation}
and
\begin{equation} \label{eq:numsat}
n_{{\rm sat}} = \int_0^\infty \langle N_{{\rm sat}} \rangle \Phi(M_h) dM_h,
\end{equation}
where
\begin{equation} \label{eq:numtot}
n = n_{{\rm cen}} + n_{{\rm sat}}.
\end{equation}

The parameter $M_{\rm min}$ primarily sets the number density for the centrals and $M_1$ for the satellites. Since we have removed one degree of freedom by holding the total number density fixed, we are free to combine these two parameters into a single parameter: $M_1/M_{\rm min}$. 
This ratio is directly influenced by the ratio of central to satellite dark matter halos predicted by dark matter simulations. Since this is decided by our cosmological prior, we choose to hold constant the $M_1/M_{\rm min}$ parameter measured from the UniverseMachine. Then, once a value is chosen for each free parameter, we can individually derive $M_1$ and $M_{\rm min}$ by numerically inverting Equation~\ref{eq:numtot}.

We remove another degree of freedom in our model by holding the fiducial value of $M_0$ fixed. This is common practice because the observables that we examine are not sensitive to large changes in this parameter.
Therefore, we only tune two free parameters in our conservative HOD model: $\sigma$ and $\alpha$. We demonstrate the effect these parameters have on $w_{\rm p}(r_{\rm p})$ predictions in Figure~\ref{fig:2pcf}. Increasing $\sigma$ decreases clustering at all scales, while increasing $\alpha$ increases clustering, especially at small scales. Pushing to smaller-scale measurements will therefore be greatly beneficial in breaking the degeneracy of these parameters.

\section{Constraints on the HOD}
\label{sec:constraints}

Following the standard methodology, we constrain the HOD by empirical measurements of number density and clustering of the galaxy population. Therefore, in this section, we present mock measurements of the stellar mass function and the projected two-point correlation function.

\subsection{Stellar Mass Function}
\label{sec:smf}

The stellar mass function (SMF), $\Phi(M_\ast)$, measures the number density of a galaxy population subdivided into bins of stellar mass. This is one of the most direct measurements of the efficiency of galaxy evolution, tracking the overall growth of galaxies over cosmic times from the accumulation of star formation. In terms of the galaxy-halo connection, the SMF provides an estimate for the SHMR, if we assume a good rank-order correlation between stellar and halo mass, as is done by abundance matching models.

\begin{figure}[ht!]
\includegraphics[width=0.47\textwidth]{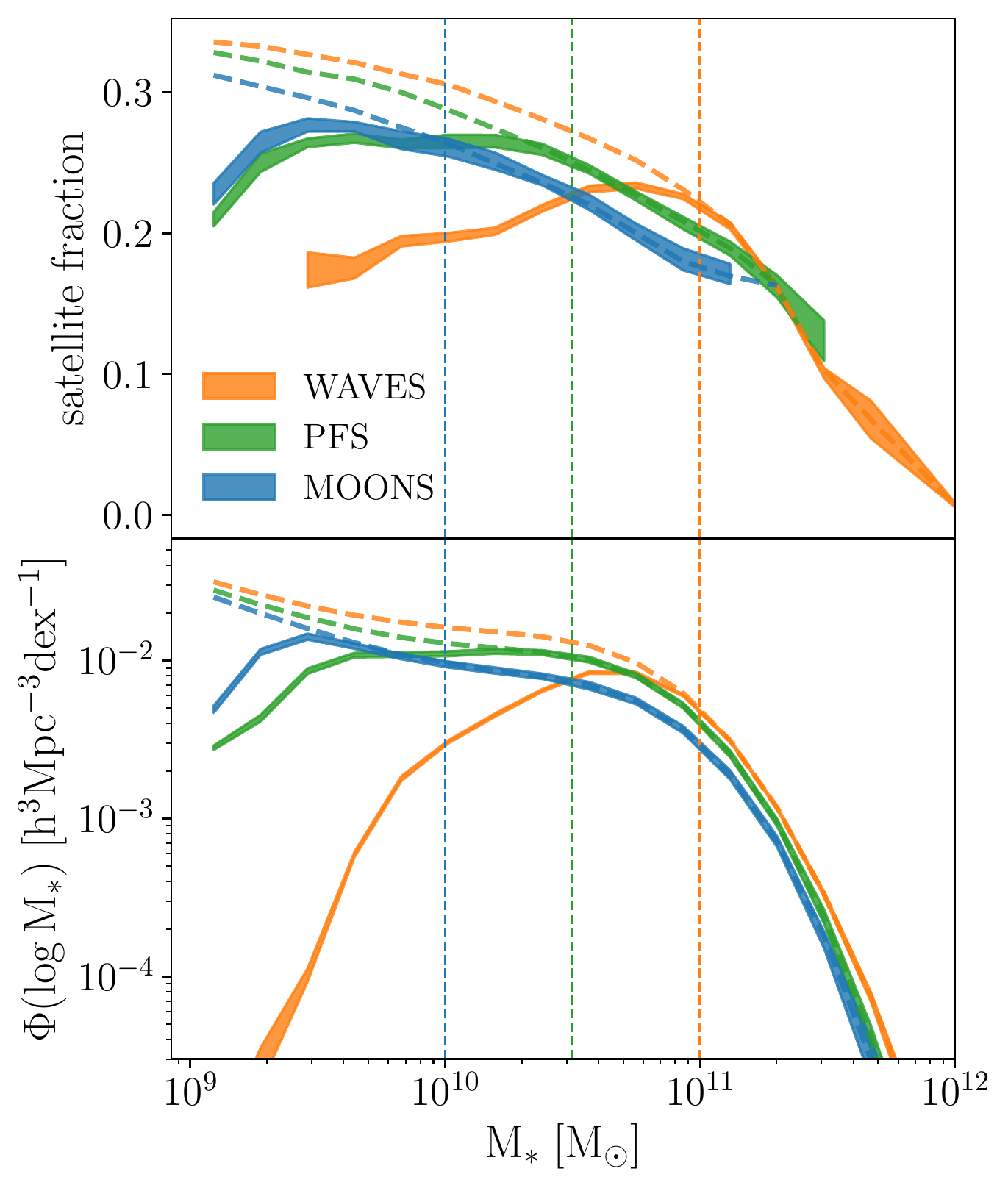}
\caption{Mock measurements of the stellar mass function (bottom) and satellite fraction (top). We plot WAVES ($0.5 < z < 0.8$) in orange, PFS ($0.8 < z < 1.2$) in green, and MOONS ($1.2 < z < 1.6$) in blue. Each band represents the 1$\sigma$ confidence region of each mock measurement, inferred by independent realizations (note that the satellite fraction here assumes perfect central/satellite assignment). Vertical dashed lines represent the stellar mass threshold we impose on each survey, while the thick dashed curves represent the true functions without photometric target selection. The observed functions agree very well with the true functions above the respective mass thresholds.
\label{fig:smf}}
\end{figure}

Spectroscopic surveys like PFS, WAVES, and MOONS will improve stellar mass estimates of galaxies from the respective epochs they are probing due to the significantly increased precision via spectroscopic redshifts, as well as tighter constraints on mass-to-light ratios from stellar ages obtained by stellar population synthesis. This will substantially improve our certainty on the distribution of stellar masses \citep{Muzzin:2009}, while abundance measurements will be further solidified by larger photometric surveys. Therefore, from this combination of data sources, we expect tight constraints on the SMF for a wide range of redshifts, especially at $z > 1$, in the coming years.

In each of our redshift samples, we measure the SMF over an ensemble of 25 mock survey realizations to quantify the uncertainty of a single survey. We present the mock SMFs of each survey in the bottom panel of Figure~\ref{fig:smf}. Note that each mock SMF agrees with the truth down to the indicated mass threshold, which is a good sign that the survey samples will be representative of the true galaxy populations. Additionally, the upper panel of Figure~\ref{fig:smf} shows the satellite fraction as a function of stellar mass, which is also in good agreement down to the completeness limit for each sample.

As discussed in Section~\ref{sec:selections}, we are able to quantify the mass completeness of each sample by selecting all mock galaxies over the mass threshold and calculating the fraction of them which are under the survey's magnitude limits. Over the entire redshift range, each sample is well over 99\% complete down to its mass threshold (see Table~\ref{tab:galaxy-samples}).

Note that the SMF is typically used as a direct constraint for the HOD through Equation~\ref{eq:smf}. However, in our conservative model, the SMF is used as a hard prior because we do not allow the total number density of galaxies to vary (see Section~\ref{sec:conservative-hod}).

\subsection{Two-Point Correlation Function}
\label{sec:2pcf}

\begin{figure*}[ht!]
\includegraphics[width=\textwidth]{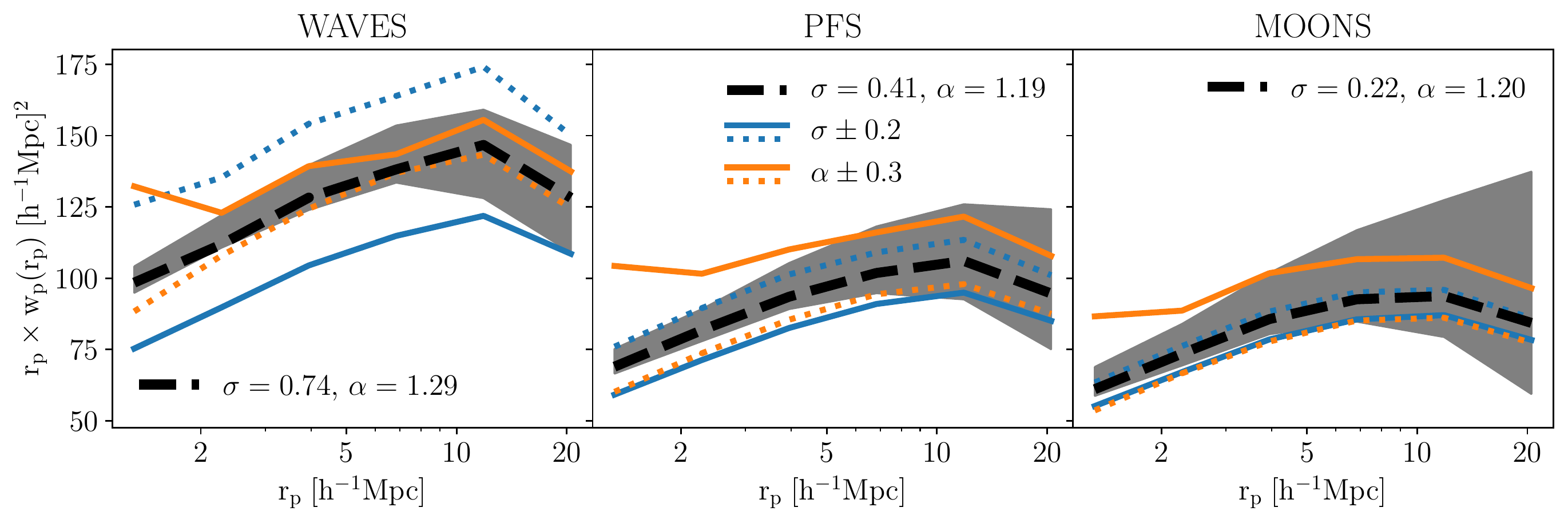}
\caption{The projected two-point correlation function. Mock $1\sigma$ constraints from each survey are given by grey shaded regions. The UniverseMachine ``truth'' HOD model is represented by the thick black dashed line, while the colored solid and dotted lines show the effect of increasing or decreasing one parameter at a time, respectively. High-mass samples like WAVES are much more sensitive to changes in $\sigma$, and low-mass samples like MOONS are more sensitive to changes in $\alpha$. These parameters produce similar observational effects, but this degeneracy can be broken by probing smaller scales.
\label{fig:2pcf}}
\end{figure*}

The two-point correlation function, $\xi(r)$, is a canonical constraint on the HOD because it measures the clustering strength of the galaxy population, which is indicative of the clustering strength of the underlying halo population. Since halo clustering is a strong function of halo mass, the two-point correlation function is very sensitive to the typical mass of the halo population \citep{Zehavi:2005, Reddick:2013, Wang:2019}.
We adopt the projected correlation function, which is defined as
\begin{equation} \label{eq:wp}
    w_{\rm p}(r_{\rm p}) = 2 \int_0^{\pi_{\rm max}}\xi(r_{\rm p}, \pi) d\pi,
\end{equation}
where we choose $\pi_{\rm max} = 50~h^{-1}~{\rm Mpc}$. The projected correlation function conveniently integrates out most of the dependence on redshift-space distortions. This is desired because our galaxy-halo connection model has no dependence on velocity dispersion, and we do not want our observable to be sensitive to that level of detail.

We perform this computation in six $r_{\rm p}$ bins with logarithmically spaced edges from 1-27~$h^{-1}~{\rm Mpc}$.
Using a relatively small number of bins here helps reduce the number of realizations needed to calculate the covariance matrix, which must be very precise for our analysis.
Due to the systematic sampling caused by fiber collisions in multiplexed spectroscopic surveys, we don't attempt to calculate the two-point correlation function below a scale of 1~$h^{-1}$~Mpc (i.e., 106 arcsec at $z=0.8$, 78.7 arcsec at $z=1.2$, and 65.2 arcsec at $z=1.6$). 
The fiber positioner patrol diameters of the instruments used by WAVES, PFS, and MOONS will likely be similarly sized, so fiber collisions should not dominate our uncertainty and the effect will be mostly mitigated by revisiting fields multiple times.

We calculate the projected two-point correlation function using the \citet{Landy:Szalay:1993} estimator implemented in the \verb|DDrppi|, \verb|DDrppi_mocks|, and \verb|convert_rp_pi_counts_to_wp| functions from the \verb|Corrfunc| package \citep{Sinha:2020}. We measure $w_{\rm p}(r_{\rm p})$ from each mock survey in Figure~\ref{fig:2pcf}, and compare it to various predictions of our conservative HOD model (see Section~\ref{sec:conservative-hod}). These measurements are sensitive to fairly small variations in $\sigma$ or $\alpha$, as can be seen in this figure. However, note that typically higher mass samples (e.g., WAVES) are less sensitive to $\alpha$ and lower mass samples (e.g., MOONS) are less sensitive to $\sigma$. This is because halo bias increases more rapidly as a function of mass at higher masses than lower masses, causing variations in high mass halo occupation to be more sensitive to the two-point correlation function. Conversely, at lower masses, there is less sensitivity to clustering signals except by varying the number of satellites, which dominate the two-point correlation function.

Our model predictions for the projected correlation function are calculated using a periodic box, for which we use the Bolshoi-Planck snapshot whose redshift is closest to the effective redshift for the given sample, as listed in Table~\ref{tab:galaxy-samples}. We weight our effective redshift (Equation~\ref{eq:zeff}) by pair counts, which scales with number times number density (Equation~\ref{eq:paircountweight}).
\begin{equation} \label{eq:zeff}
    z_{\rm eff} = \frac{\int_{z_{\rm min}}^{z_{\rm max}} z W(z) dz}{\int_{z_{\rm min}}^{z_{\rm max}} W(z) dz}
\end{equation}
where
\begin{equation} \label{eq:paircountweight}
    W(z) = (dN / dz) n = \frac{(dN/dz)^2}{dV/dz}
\end{equation}

\subsection{MCMC Fits}
\label{sec:mcmc}

We perform the measurement of $w_{\rm p}(r_{\rm p})$ (see Section~\ref{sec:2pcf}) on 600 independent realizations of each mock survey, seeded by randomized orientations and origins in the Bolshoi-Planck box, using the \verb|lightcone| code provided in the UniverseMachine package. We then calculate the mean and covariance matrix from these samples and define a six-dimensional multivariate normal likelihood distribution for our six $r_{\rm p}$ bins of $w_{\rm p}$. We then use the \verb|emcee| package to sample the posterior probability distribution of our HOD parameter-space: \{$\sigma$, $\alpha$\}, with a uniform prior confined to $10^{-5} < \sigma < 5$ and $0.1 < \alpha < 3$. We initialize our MCMC chains very close to the corresponding fiducial parameters given in Table~\ref{tab:hodparams}, but allow them to run many autocorrelation lengths to ensure they are well converged, as discussed in Section~\ref{sec:forecasts}.

Note that the $w_{\rm p}(r_{\rm p})$ measured by a survey could be obtained more realistically by drawing one of the 600 realizations, rather than using the mean. However, from tests of additional MCMC runs, we confirm that there is no strong bias in the constraining power from using individual realizations instead of using the mean value. Therefore, we define our likelihood using the mean, which is more stable and the only fair comparison between measurements using various survey parameters. Note that cosmic variance and measurement error is still incorporated through the covariance matrix.

\begin{figure*}[ht!]
    \includegraphics[width=0.33\textwidth]{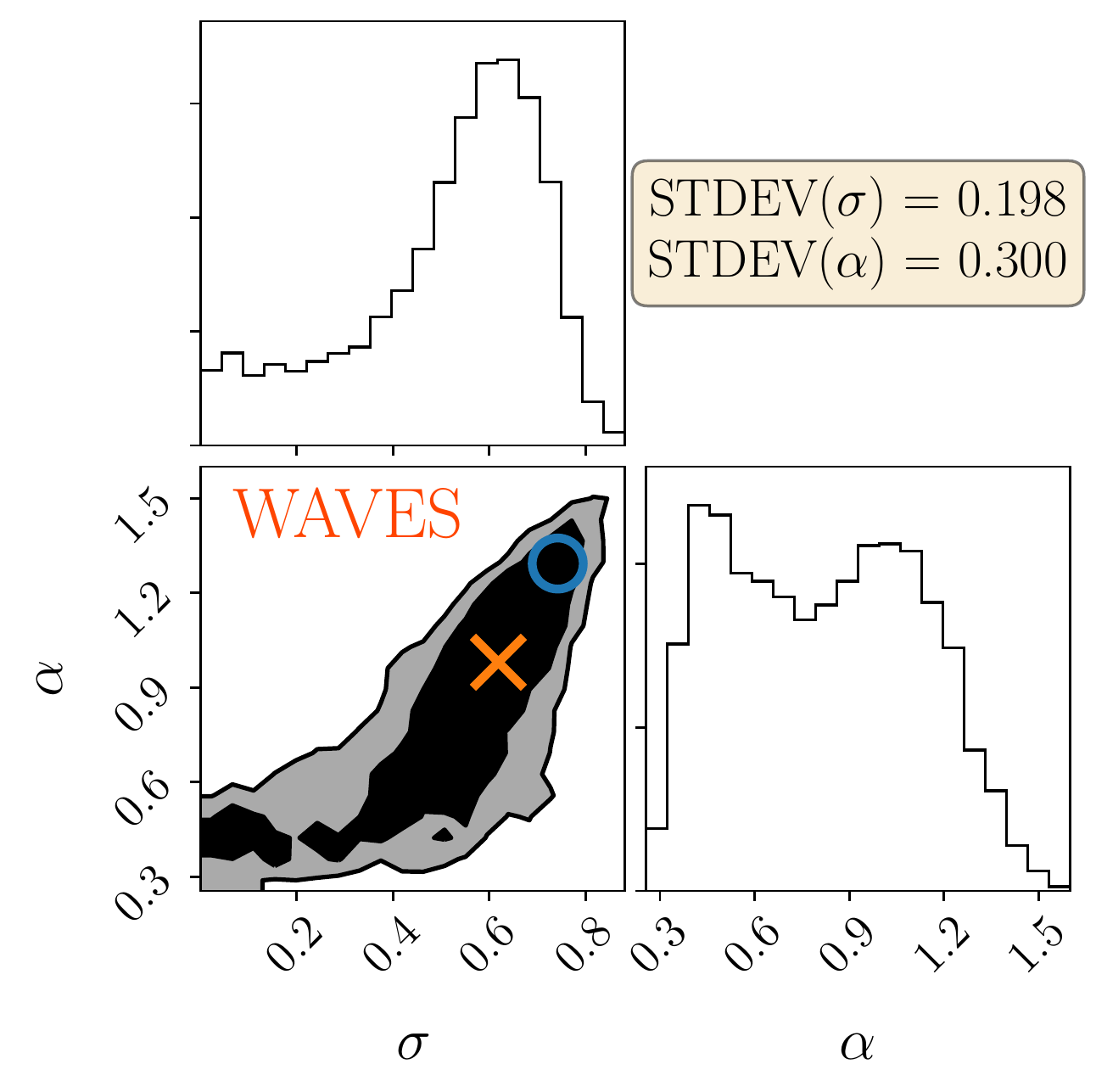}
    \includegraphics[width=0.33\textwidth]{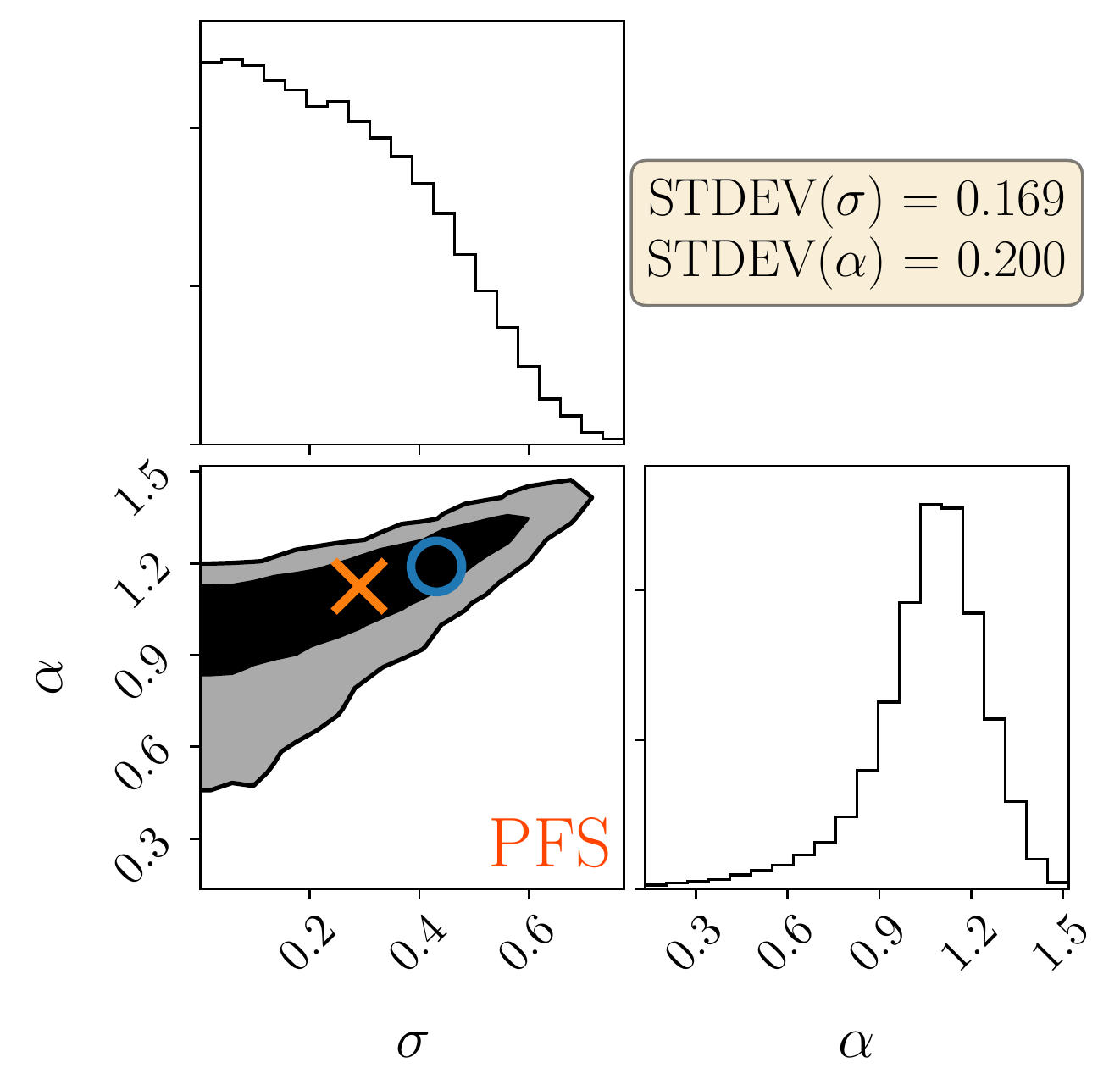}
    \includegraphics[width=0.33\textwidth]{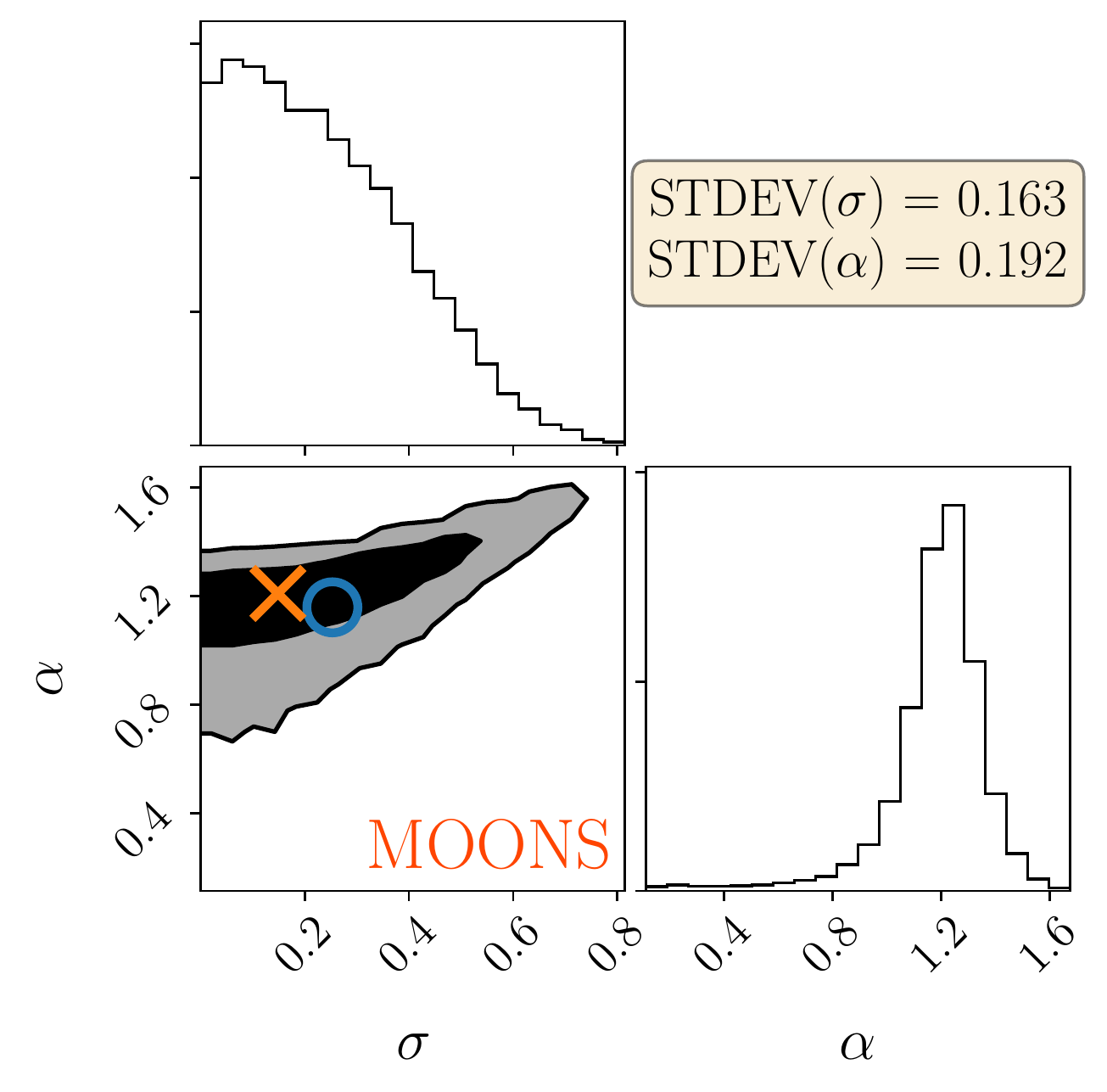}
    \caption{HOD posterior probability distribution measured in WAVES (left), PFS (center), and MOONS (right). Measured by MCMC sampling of our two-parameter conservative HOD model at the effective redshift of each sample, and comparing the predicted $w_{\rm p}(r_{\rm p})$ to 600 mock realizations. For each sample, the UniverseMachine truth value is marked with a blue circle and best-fit parameters are marked with an orange X. \label{fig:mcmc}}
\end{figure*}

\section{Results: Predictions for Next-Generation Surveys}
\label{sec:analysis}

\subsection{Forecasts for WAVES, PFS, and MOONS}
\label{sec:forecasts}

Thanks to the small number of free parameters in our model, we obtain favorably high acceptance rates ($\sim 50\%$) and low autocorrelation lengths ($\sim 50$), which helps reduce the time required to run our MCMC chains.
To ensure we are not biased by our initial guess, we removed a burn-in of 250 iterations from the beginning of each chain, although this has a very small effect due to the long length of our chains.
In each MCMC, we sample 150,{}000 trial points, yielding posteriors of very high resolution. We present a corner plot of the posterior measured in each mock survey in Figure~\ref{fig:mcmc}. These posteriors show that our method does a good job of constraining our HOD to a fairly small region in parameter space. The largest difficulty is constraining the $\alpha$ parameter in the WAVES sample due to the high mass centrals dominating both the number density and clustering signal, and a large covariance between $\sigma$ and $\alpha$.

\begin{figure}[ht!]
\includegraphics[width=0.47\textwidth]{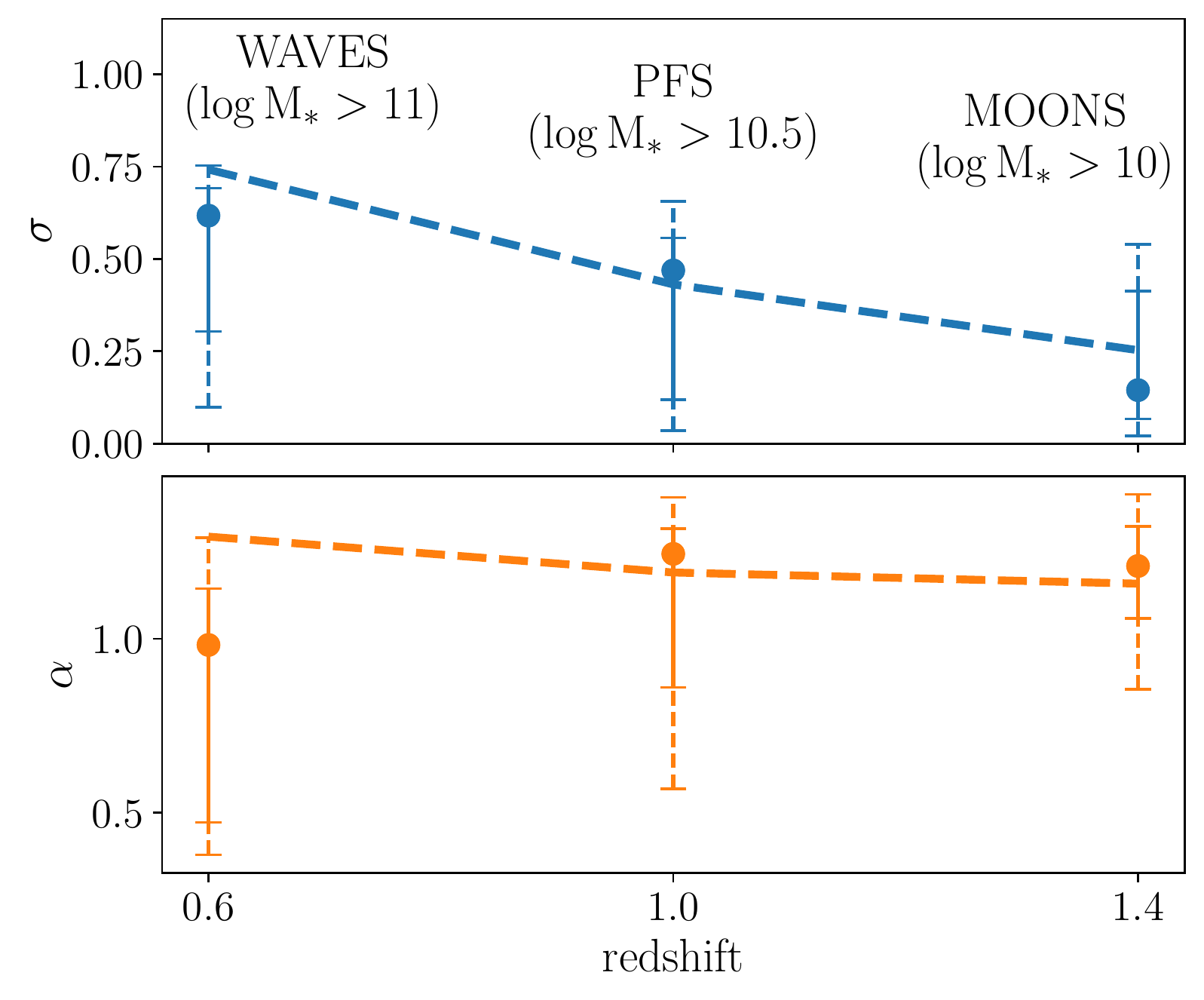}
\caption{Predicted constraints on the evolution of the HOD. We compile best fit measurements along with 5th, 16th, 84th, and 95th percentile of the posterior (see Figure~\ref{fig:mcmc}) of $\sigma$ (upper panel) and $\alpha$ (lower panel). UniverseMachine ``truth'' values are connected by dashed lines.
\label{fig:hod-vs-sample}}
\end{figure}

Putting together the information from the posteriors from WAVES, PFS, and MOONS, we will be able to constrain the HOD across a wide range of mass and redshift. We compile the predicted constraints on the evolution of these HOD parameters in Figure~\ref{fig:hod-vs-sample}. Certain parameters will still be very poorly constrained using this type of analysis; for example, $\alpha$ in the WAVES sample. This is primarily because the two-point correlation function is most sensitive to $\sigma$ at high masses and $\alpha$ at low masses, but additional metrics may be able to provide more information (see Appendix~\ref{sec:additional-metrics}).

Note that the WAVES sample produces very poor constraints on $\alpha$ (the satellite occupation slope parameter), whereas the MOONS sample produces particularly strong constraints on $\alpha$. This demonstrates a fundamental difficulty of using constraints only from number density and the two-point correlation function. Since the two-point correlation function is primarily sensitive to the most clustered data, it is more informative for satellites at lower mass thresholds and centrals at high mass thresholds (as seen in Figure~\ref{fig:2pcf}). However, for the WAVES sample, note that $\sigma$ and $\alpha$ are nearly degenerate, which results in relatively poor constraints for both parameters.

It should be noted that improving our techniques may lead to tighter constraints on these regions of difficulty. First of all, it is possible to partially break the degeneracy between central and satellite clustering by measuring the two-point correlation function to smaller scales (sub-Mpc) using fiber collision corrections. Additionally, alternative statistics like counts-in-cylinders tend to be more sensitive to certain galaxy populations, and therefore provide excellent complementary information to the two-point correlation function \citep{Wang:2019}.
This will be important for analyzing the real data, but will come at a greater computational cost, particularly because this increases the size of the covariance matrix, and will likely require many more mock realizations to calculate accurately.

\subsection{Measurement Error vs. Survey Parameters}

Telescope time is typically the limiting factor in survey design. To first order, the amount of time a survey requires is roughly proportional to the number of objects observed. It is therefore possible to either scale up the sky area in exchange for a decrease in {targeting} completeness or vice versa. Increasing the sky area of a survey would increase the volume and therefore decrease the cosmic sample variance; on the other hand, high {targeting} completeness helps mitigate the uncertainty of small-scale pair counts, especially when accounting for fiber collisions \citep{Bianchi:Percival:2017}. Although this is not tested in this work, it should also be noted that higher {targeting} completeness should be favored if the goal is to increase the accuracy of identifying central galaxies in group reconstruction \citep{Looser:2021}.

\begin{figure*}[ht!]
\includegraphics[width=\textwidth]{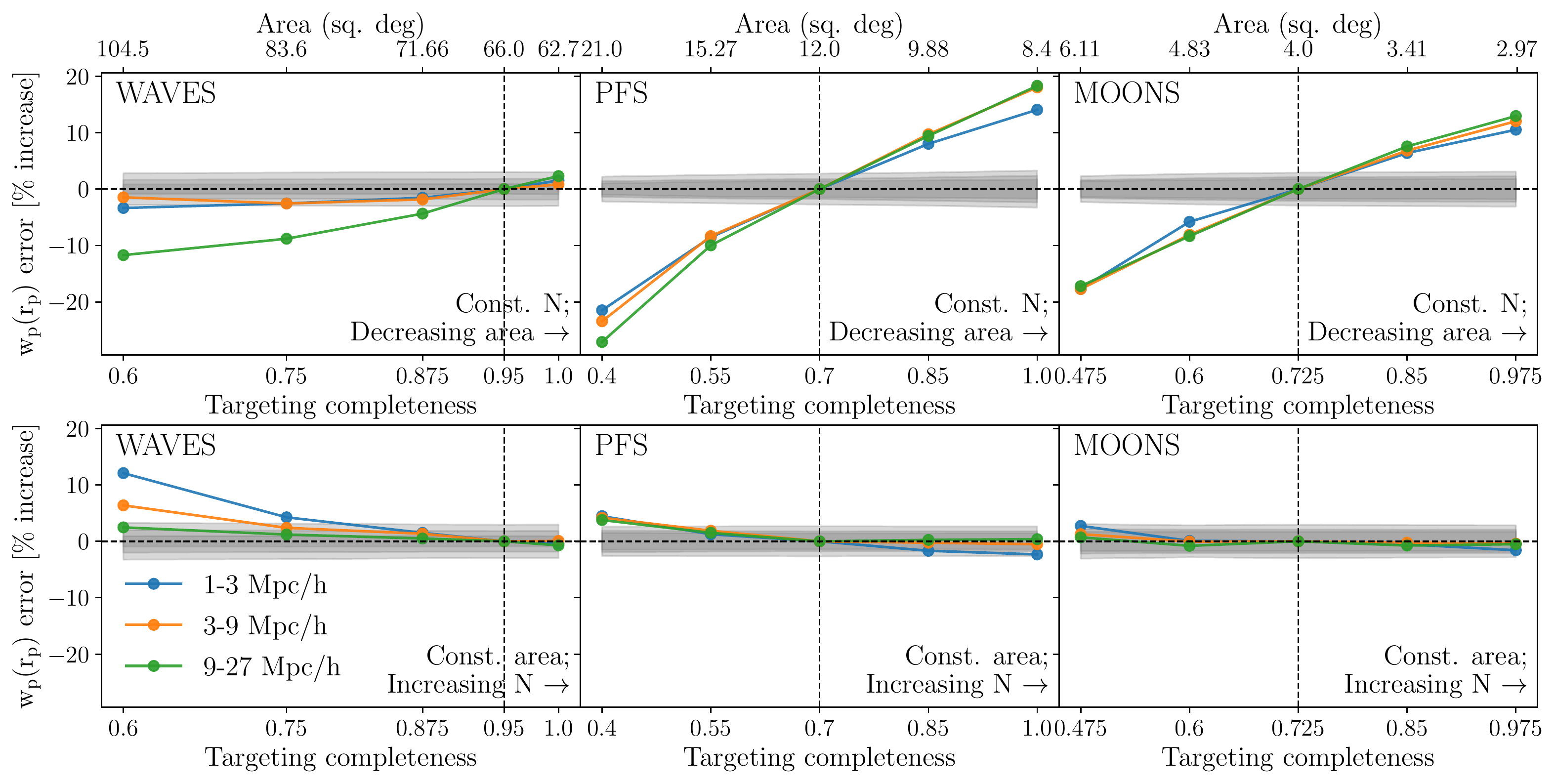}
\caption{Precision of mock $w_{\rm p}$ measurements, color-coded by $r_{\rm p}$ scale, as a function of {targeting} completeness. Characteristic jackknife uncertainties are shown with grey bands. In the top panels, the sky area is varied to conserve the total number of targets (given in Table~\ref{tab:galaxy-samples}). In the bottom panels, the sky area is conserved at the true value for each survey, and therefore the number of targets increases with {targeting} completeness. Particularly for the PFS and MOONS samples, there are significantly stronger trends in the top panels. This suggests that the uncertainties in $w_{\rm p}(r_{\rm p})$ are dominated by cosmic variance, as opposed to shot noise.
\label{fig:survey-wp-err}}
\end{figure*}

Using our mock catalogs to estimate uncertainties, we vary these survey parameters to quantify their effects on constraining power. In Figure~\ref{fig:survey-wp-err}, we present the dependence of survey parameters on the uncertainty of the projected two-point correlation function. For PFS- and MOONS-like surveys, the only way to greatly reduce uncertainties is by increasing the area (see top row). Increasing the number of targets in the same area makes much more modest improvements (see bottom row). In other words, smaller area surveys like PFS and MOONS are dominated by cosmic variance, not shot noise. Wider surveys like WAVES (left panel) typically find that the two-point correlation function uncertainty is dominated by cosmic variance on large scales (9-27~$h^{-1}$~Mpc) and shot noise on small scales (1-3~$h^{-1}$~Mpc).

To roughly quantify the effect survey parameters have on correlation function uncertainties, we calculate the percent decrease in $w_{\rm p}$ error at $r_{\rm p}$ = 3-9~$h^{-1}$~Mpc from the true survey parameters for two cases: (1) {5\% increased} area and {$\sim$5\% decreased targeting} completeness (conserving the number of targets) and (2) {5\% increased} {targeting} completeness and the same area. {We interpolate these numbers with} linear regression to fit the slope of the orange lines in Figure~\ref{fig:survey-wp-err}. We find that, {for a fixed number of targets, a 5\% increased} survey area decreases the uncertainty of the correlation function at intermediate scales by {0.15\%, 1.2\%, and 1.1\%} for our WAVES, PFS, and MOONS samples, respectively. For a fixed survey area, {a 5\% increase in} the number of targets improves the same constraints by {0.7\%, 0.25\%, and 0.1\%}.

{It should be noted that increasing targeting completeness of a multiplexed survey does not necessarily increase linearly with telescope time, particularly as the targeting completeness approaches 100\%. Even though we do not explicitly run targeting simulations to connect our survey parameters to telescope time, our comparison in Figure~\ref{fig:survey-wp-err} is still useful to test the effective significance of shot noise vs.\ cosmic variance. The results of this comparison indicate} that the samples probed by PFS and MOONS are dominated by cosmic variance and can only be improved significantly by increasing the observing area.

\begin{figure*}[ht!]
\includegraphics[width=\textwidth]{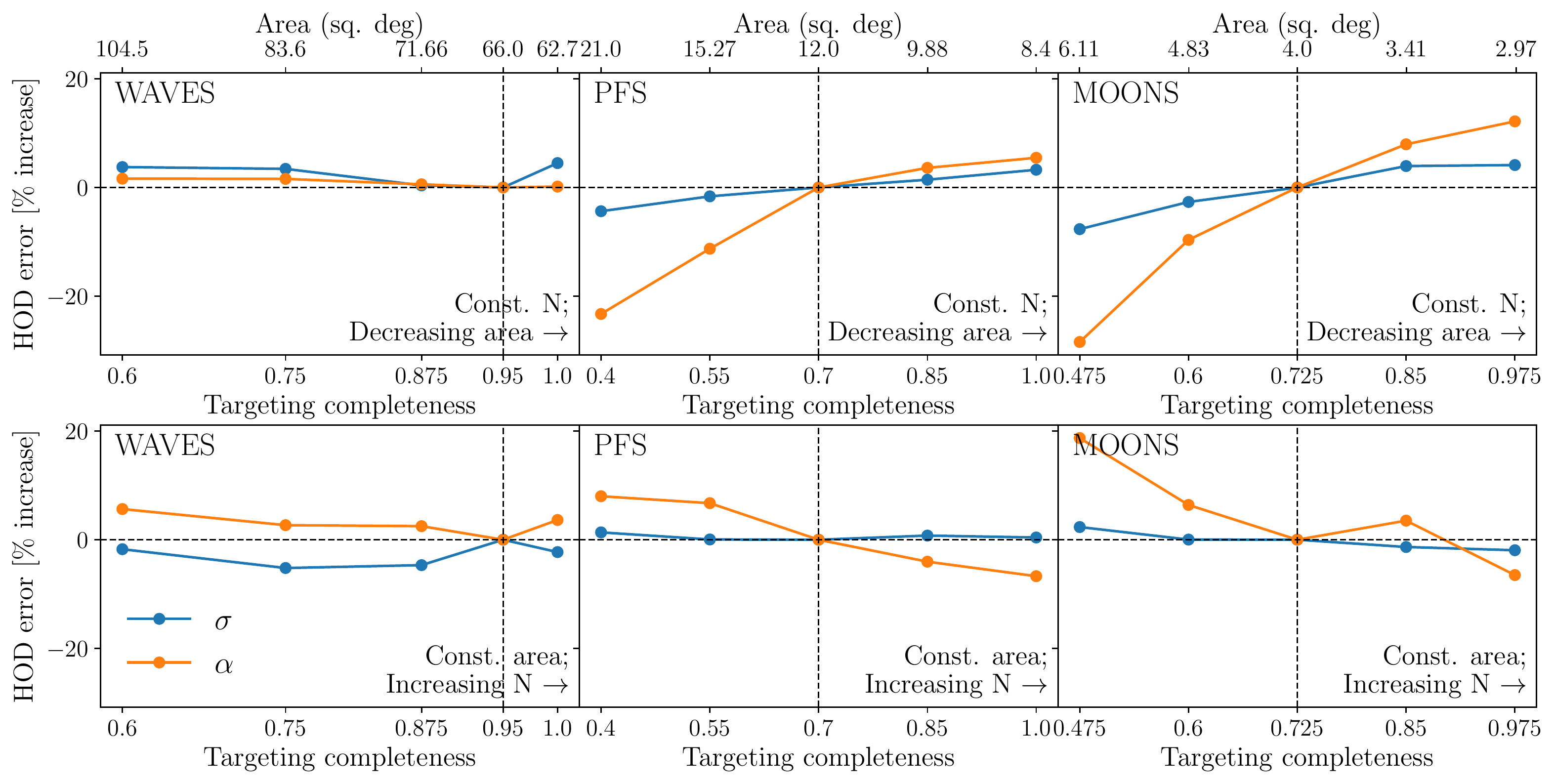}
\caption{Precision of HOD parameters $\sigma$ (blue) and $\alpha$ (orange), as a function of {targeting} completeness. The panels are arranged analogously to Figure~\ref{fig:survey-wp-err}. These uncertainties do not appear to be dominated by cosmic variance as strongly as the two-point correlation function. They are affected just as strongly by shot noise. This demonstrates the importance of propagating uncertainties from the full covariance matrix, rather than just the diagonal components shown in Figure~\ref{fig:survey-wp-err}.
\label{fig:survey-hod-err}}
\end{figure*}

Given the covariance matrix of $w_{\rm p}(r_{\rm p})$ calculated for each set of survey parameters, we also calculate HOD constraints via our MCMC method. We present the HOD constraining power as a function of survey parameters by varying {targeting} completeness and area in Figure~\ref{fig:survey-hod-err}.
In the WAVES survey, we only find very small changes in constraints as we vary the survey parameters.
For PFS and MOONS, the difference between Figures~\ref{fig:survey-hod-err} and~\ref{fig:survey-wp-err} indicates that the covariance of the two-point correlation function with respect to various regions of the universe is significantly different from the covariance due to varying HOD parameters. It appears that this allows the HOD to be somewhat more robust to cosmic variance than $w_{\rm p}$. Throughout, the constraints on $\alpha$ may be slightly more dependent on survey parameters than the constraints on $\sigma$.

\subsection{Comparisons to Past Surveys}

Surveys like WAVES, PFS, and MOONS will be monumental in pushing measurements of the two-point correlation function to higher redshifts because they will provide the precise spectroscopic measurements necessary to perform those calculations. There are currently no existing spectroscopic datasets that are comparable in size to the $z > 1$ samples we will obtain from PFS and MOONS.

For the slightly lower redshifts probed by WAVES, the closest comparison would be prism surveys such as the PRIsm MUlti-object Survey \citep[PRIMUS;][]{Coil:2011} or the Carnegie-Spitzer-IMACS \citep[CSI;][]{Kelson:2014}. PRIMUS, the larger of these two surveys, has measured the spectra of galaxies out to $z \sim 1.2$ with a redshift precision of $\sigma_z / (1+z) \sim 0.005$. Incorporating all fields that overlap with imaging from the Galaxy Evolution Explorer \citep[GALEX;][]{Martin:2005}, Spitzer Space Telescope \citep{Werner:2004}, Infrared Array Camera \citep[IRAC;][]{Fazio:2004}, and various ground-based surveys, PRIMUS has an area of 5.5 deg$^2$ and is complete down to similar stellar masses as WAVES \citep[see][]{Moustakas:2013}.

\begin{figure}
    \includegraphics[width=0.47\textwidth]{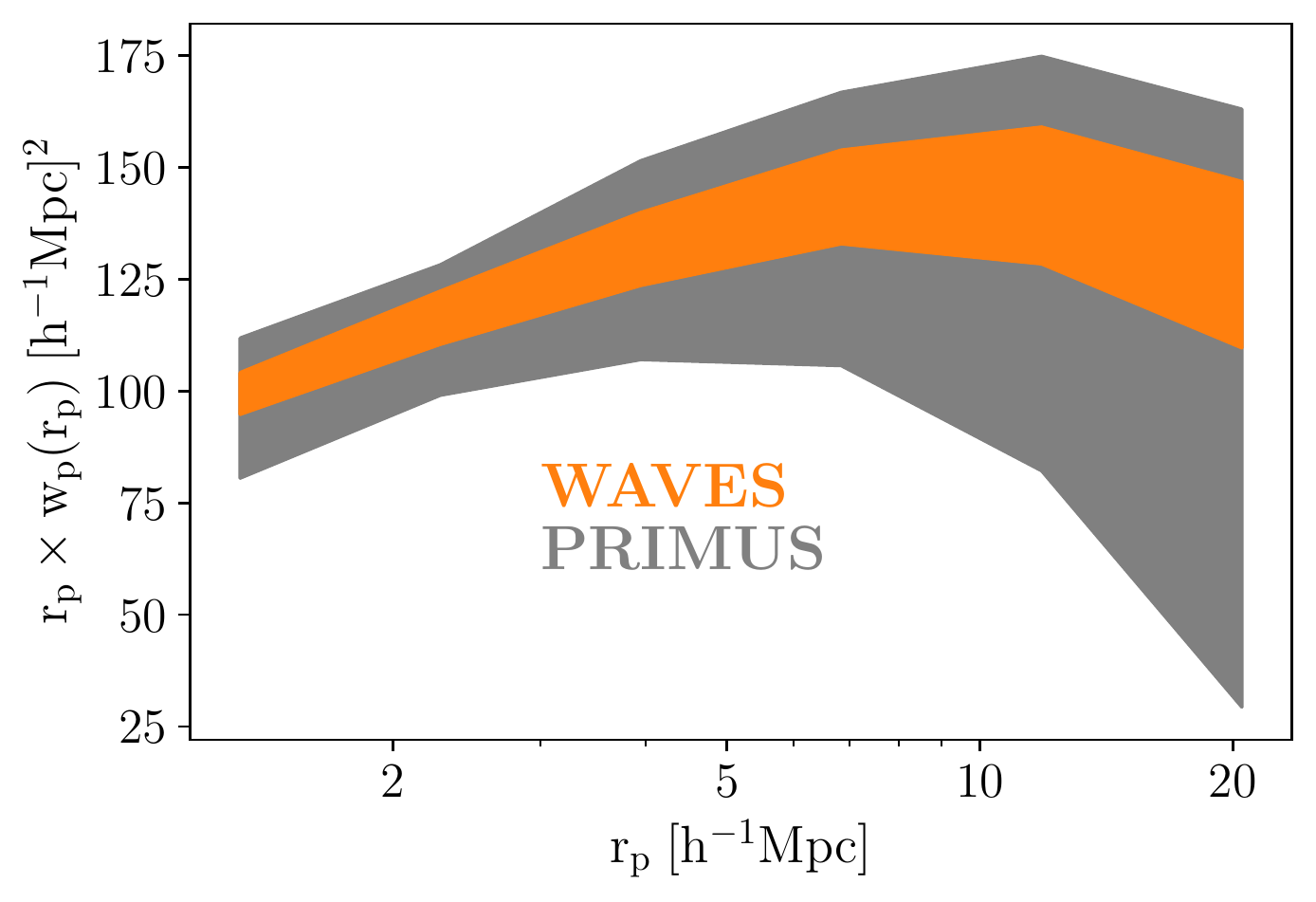}
    \caption{Mock measurements of the projected two-point correlation function for WAVES vs. PRIMUS. Note that we only used 25 independent realizations to quantify the observational error of the PRIMUS measurement, compared to the 600 used for WAVES. The increased precision from WAVES is due to its area (66 deg$^2$) which is over 10 times that of PRIMUS (5.5 deg$^2$). A small systematic offset is driven by redshift errors in PRIMUS, which we assume will be negligible compared to velocity distortions in WAVES.
\label{fig:wp-waves-vs-primus}}
\end{figure}

We compare mock measurements of the projected two-point correlation function of our WAVES galaxy sample ($0.5 < z < 0.8$ and $\log(M_\ast / M_\odot) > 11$) for the survey parameters of WAVES and PRIMUS in Figure~\ref{fig:wp-waves-vs-primus}. This plot illustrates the significantly increased precision we can expect to obtain from this sample of galaxies. Additionally, unlike PRIMUS, we assume that the redshift uncertainties in WAVES will be negligible compared to redshift distortions. The additional redshift error from PRIMUS does not greatly contribute to the errorbars of the two-point correlation function, but this does produce a small systematic offset which may further reduce the correlation function's sensitivity to HOD parameters.

\begin{figure}
    \includegraphics[width=0.47\textwidth]{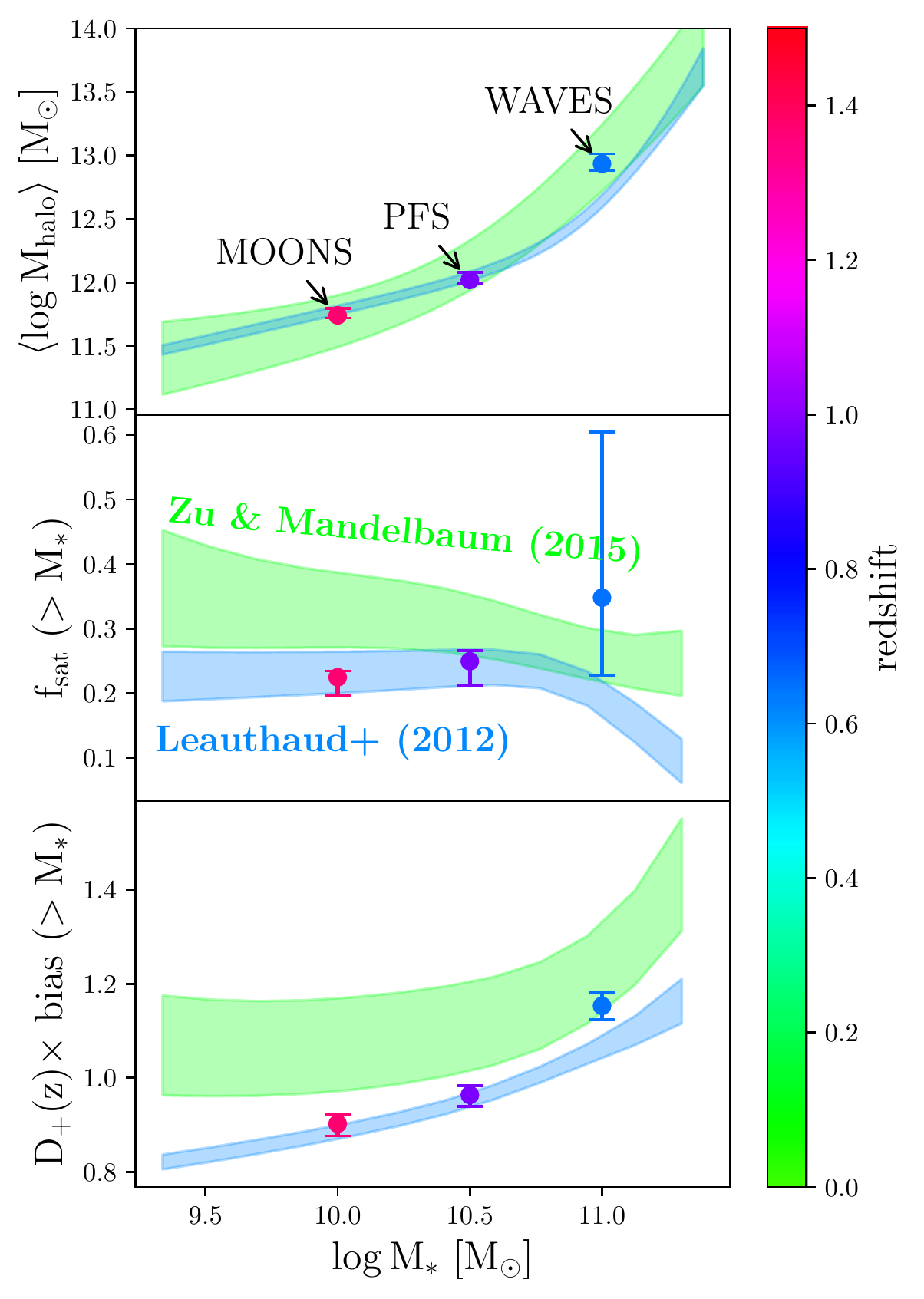}
    \caption{Compiled measurements of the galaxy-halo connection. As a function of stellar mass, we show several studies that have measured the mean halo mass (top panel), satellite fraction above the stellar mass threshold (middle panel), and absolute bias above the stellar mass threshold (bottom panel). Projected HOD analysis from this work for WAVES, PFS, and MOONS is shown by the colored points, which push to significantly higher redshifts with comparable uncertainties to previous studies.
    \label{fig:compare-to-literature}}
\end{figure}

However, most of our current understanding of the galaxy-halo connection comes from studies of surveys that either span lower redshifts or rely on photometric redshifts. For example, \citet{Zu:Mandelbaum:2015} use photometric redshifts from SDSS \citep{Blanton:2017} and \citet{Leauthaud:2012} from COSMOS \citep{Scoville:2007}. In Figure~\ref{fig:compare-to-literature}, we present key findings of these past studies in terms of the SHMR, absolute bias, and satellite fraction of several stellar mass threshold galaxy samples and compare to the same measurements derived from our HOD projections of the WAVES, PFS, and MOONS surveys. These upcoming surveys will push to significantly deeper redshifts than past surveys, with comparable uncertainties. 

Measuring the HOD for various surveys at distinct redshifts and stellar mass thresholds has been and will continue to be a powerful tool for studying galaxy evolution as measured by mean halo mass, satellite fraction, and galaxy bias (see Figure~\ref{fig:compare-to-literature}). Currently, we have little evidence for significant redshift evolution in most of these properties. However, more precise and higher redshift measurements of these parameters will give us a much clearer picture of how they may evolve with the age of the universe. Each of these metrics is highly sensitive to various models of galaxy formation, so these new measurements will have a large impact on how we think about the important processes driving the shut down of rapid star formation that occurred at cosmic noon.

\section{Conclusions}
\label{sec:conclusions}

In this paper, we present the CLIMBER procedure, which we use to calibrate photometry into the UniversemMachine and similar models. This procedure performs well at reproducing a broad range of properties simultaneously. Twenty-five realizations of the $0.7 < z < 1.7$ mock catalog used for the PFS and MOONS samples in this work are available at \url{https://alanpearl.github.io/\#data}. Alternatively, you may install the utilities we used to construct these catalogs at \url{https://github.com/AlanPearl/mocksurvey}.

We have used our mock catalogs to test the forthcoming generation of massively multiplexed spectroscopic galaxy surveys, which will likely change our understanding of galaxy formation, the galaxy-halo connection, and possibly even cosmology in profound ways. The high-redshift samples being probed will provide new constraints on theories and models which predict how populations evolve with redshift. The UniverseMachine will face new scrutiny in its ability to reproduce clustering and environmental quenching signals in the distant universe. The constraints on parameters of interest in the UniverseMachine will likely tighten significantly to match the new observations. It may even be possible that the UniverseMachine will require reparameterization in order to achieve a good fit to the new data. Either way, analysis of these new surveys will greatly impact our understanding of the evolution of the galaxy population's connection to its dark matter environment over the history of the universe.

Using the two-point correlation function, we have found that surveys such as WAVES, PFS, and MOONS will place new constraints on the galaxy-halo connection. We characterize constraints on the central term of the HOD with the parameters $\sigma$ and $\log M_{\rm min}$. The precision to which we measure these parameters is displayed in Figure~\ref{fig:mcmc}. We have found that studies of lower mass galaxies, like the MOONS sample, will not achieve strong constraints from $w_{\rm p}(r_{\rm p})$ alone and will likely need to use other counts-in-cells statistics that are more sensitive to the weak clustering signal of low-mass centrals.

We characterize constraints on the satellite term of the HOD with the parameters $\alpha$ and $\log M_1$. The precision of these measurements is also displayed in Figure~\ref{fig:mcmc}. We find that the satellite term of the HOD will be most poorly constrained in high-mass samples, where the clustering signal is dominated by centrals. In this regime, small-scale correlation function measurements are the most sensitive to the satellite occupation of halos. Smaller-scale measurements will be possible from these surveys using fiber collision corrections, but those will likely come at the cost of large shot noise. Therefore, it is still important for follow-up surveys to improve the {targeting} completeness of these galaxy samples and reduce the effect of fiber collisions.

The achievable constraining power of these parameters is dependent on survey parameters, such as {targeting} completeness (which reduces shot noise) and area (which reduces cosmic variance). Our conclusions can be summarized by the following key points:
\begin{itemize}
    \item The two-point correlation function measurements from PFS and MOONS are both primarily dominated by cosmic variance, rather than shot noise. We have shown that, with fixed sample size, increasing their survey area drastically reduces this uncertainty.
    \item The HOD constraints from PFS and MOONS are less dominated by cosmic variance. This demonstrates the importance of using the full covariance matrix to calculate HOD constraints.
    \item From WAVES, there is a more balanced combination of shot noise, which is dominant on small scales ($1-3~h^{-1}~$Mpc) and cosmic variance, which is dominant on large scales ($9-27~h^{-1}~$Mpc). The resulting HOD constraints are not strongly affected by small changes in survey parameters.
\end{itemize}

Another important survey parameter, which has not been explored in this work, is the number of independent fields. PFS and MOONS are both planning on dividing their survey into several fields, which will slightly mitigate some of their large cosmic variance. Additionally, our predicted future constraints could be improved by supplementing the correlation function with counts-in-cylinders, which is more sensitive to the weak clustering of low-mass centrals. Simulating detailed targeting strategies may also be important for more precise optimizations of constraining power, as this is necessary to calculate pair counts corrections at very small scales due to fiber collisions. This is important for unbiased estimates of both the two-point correlation function and counts-in-cylinders.

There are sure to be many discoveries in store thanks to this new generation of surveys. This new data will be investigated from all angles of the galaxy-halo connection: empirical models, SAMs, and hydrodynamic simulations alike will be put to the test. We hope that through this combined effort and publicly available tools like our mock catalogs, we will be able to utilize this data to its full potential.

\acknowledgements
This research has made extensive use of the arXiv and NASA's Astrophysics Data System.
This research has made use of adstex (\url{https://github.com/yymao/adstex}).
We thank Peter Behroozi for giving helpful comments and distributing UniverseMachine data publicly. We also thank Aldo Rodr\'{i}guez-Puebla for helpful suggestions for future work. We thank the anonymous referee for suggestions that have improved this paper.

\software{
Halotools \citep{Hearin:2016},
Corrfunc \citep{Sinha:2020},
emcee \citep{Foreman-Mackey:2013},
corner.py \citep{Foreman-Mackey:2016},
scikit-learn \citep{Pedregosa:2012},
SciPy \citep{Virtanen:2020},
matplotlib \citep{Hunter:2007},
Astropy \citep{astropy:2018},
NumPy \citep{vanderWalt:2011},}

\bibliography{bibtex}

\begin{figure*}[ht!]
\centering
\includegraphics[width=\textwidth]{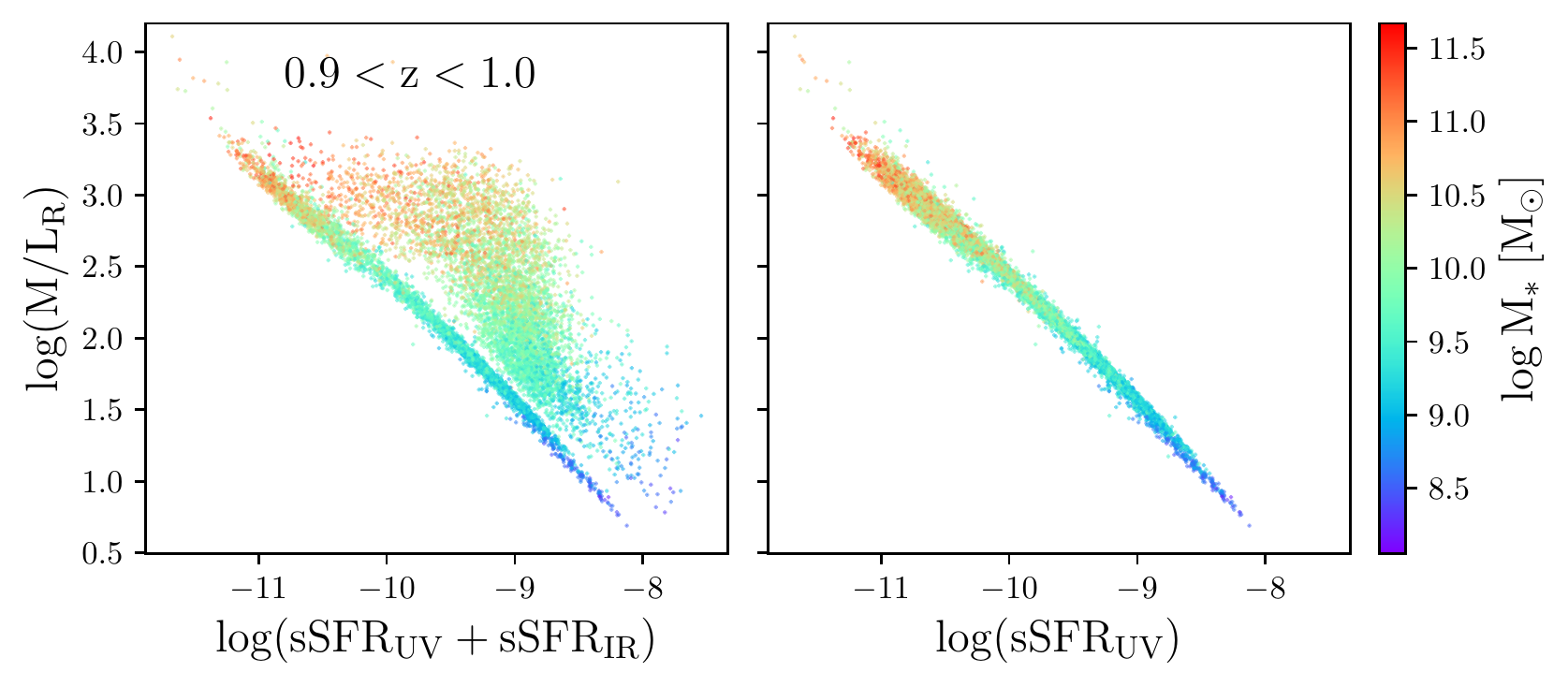}
\caption{The relation between specific star formation rate and mass-to-light ratio for UltraVISTA galaxies in the redshift slice $0.9<z<1.0$. In the left panel, we use total sSFR (UV + IR), and in the right panel, we only use UV sSFR. The two plots share in common all galaxies without any IR detection, but the galaxies with IR detection form a cloud to the right due to their higher total sSFR (removing galaxies without IR detection removes the remaining narrow distribution entirely). We choose not to include $\mathrm{sSFR_{IR}}$ in our mass-to-light calibration due to this discontinuity between detections and non-detections.
\label{fig:ssfr-uv-vs-tot}}
\end{figure*}

\appendix
\section{CLIMBER Details}
\label{sec:climber-details}

The goal of Calibrating Light: Illuminating Mocks By Empirical Relations (CLIMBER) is to estimate the luminosity of each mock galaxy in any observed photometric band. Since star-forming galaxies host more young blue stars, color is a smooth function of specific SFR (sSFR). Both of these quantities correlate strongly with the mass-to-light ratio, as shown by \citet{Bell:deJong:2001}. The relationship between mass-to-light ratio in each band and sSFR is approximately a power law that can be empirically calibrated. 

However, we first need to ensure self-consistency between the model and empirical parameters. In the model, SFRs for the star-forming population are drawn from mass-dependent distribution (the star-forming main sequence) which evolves with redshift. The star-forming main sequence is matched to empirical distributions at similar redshifts to ensure the values are physical. However, for the quiescent population, the SFRs are drawn from a non-evolving log-normal distribution centered around an sSFR of $\mathrm{10^{-11.8}\;yr^{-1}}$. Given that this is inconsistent with empirical assumptions past $z\sim0$, we adopt only their rank-ordering.
For each mock galaxy, we first map its sSFR to an empirically calibrated value of ultraviolet sSFR (\ssfruv). We choose \ssfruv\ due to its tight correlation with mass-to-light ratio and reliable measurements in both quiescent and star-forming galaxies.

It would be even better to use the total sSFR by adding infrared (IR) sSFR, but this is not possible for the UltraVISTA dataset because 53\% of our training data have no detection in the IR. This creates an artificial discontinuity between IR detections and non-detections, thereby forming two distinct populations in the sSFR-M/L plane (see Figure~\ref{fig:ssfr-uv-vs-tot}). Including this discontinuity would greatly increase the difficulty in mapping from sSFR to M/L, particularly because it would require knowing the stellar mass to know which population a galaxy is a part of (this is primarily an observational effect in which lower mass galaxies at the same sSFR are less likely to have infrared detections). Including the stellar mass as a feature in our random forest is undesirable because it could have unknown consequences when extrapolating to lower stellar masses. Due to our decision to map sSFR to \ssfruv, we warn that our mocks may underpredict the spread in the distribution of M/L at a given sSFR. However, we believe that this is the best option due to its continuity and accuracy in reproducing magnitude and color distributions.

\begin{figure}[ht!]
\includegraphics[width=0.47\textwidth]{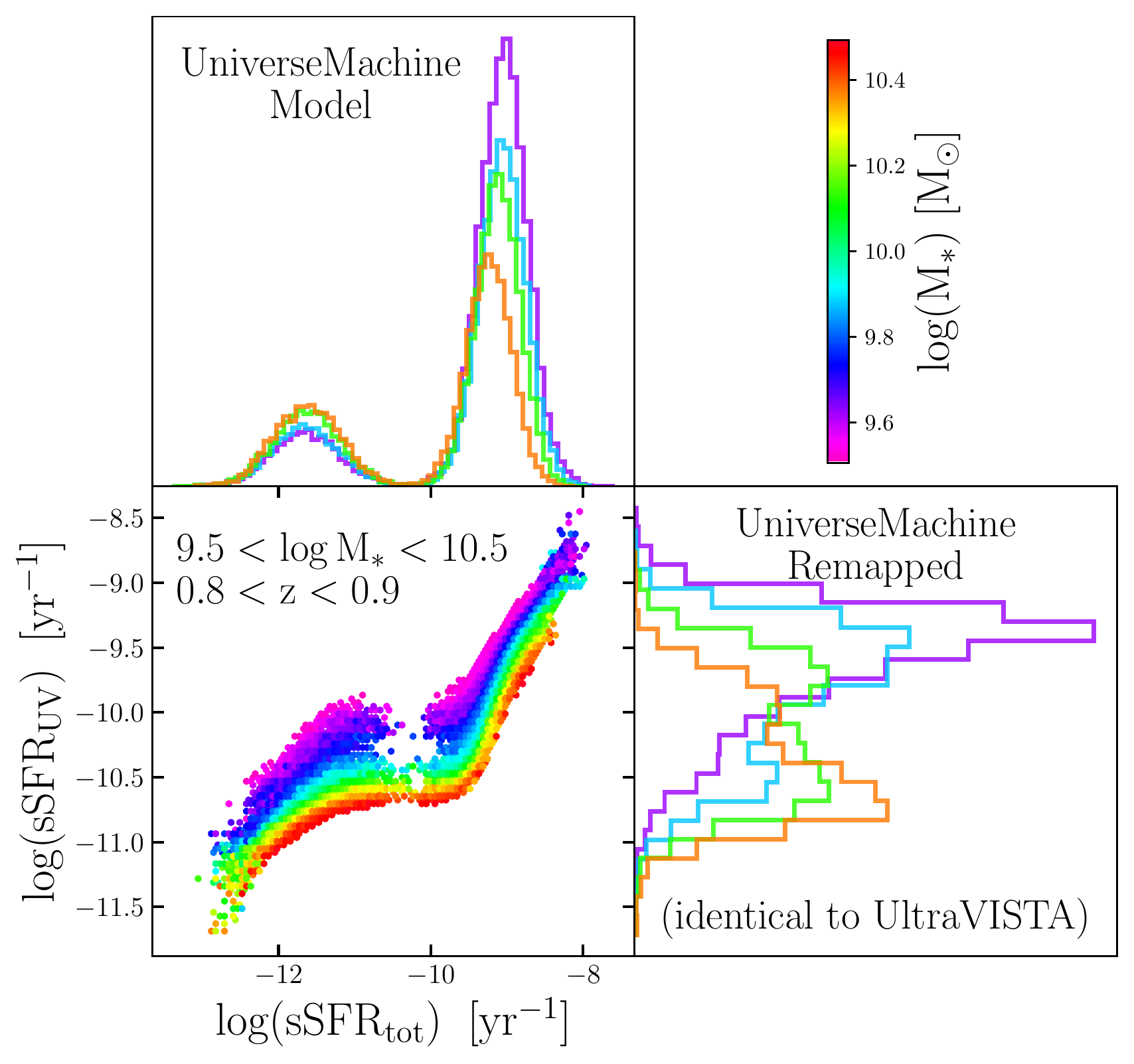}
\caption{Conditional abundance-matching mapping from $\rm sSFR \rightarrow sSFR_{UV}$ for UniverseMachine mock galaxies for a range of stellar masses at the redshift slice $0.8<z<0.9$. After being remapped, the distribution is forced to be identical to that of UltraVISTA at fixed stellar mass and redshift. Note that there is a near one-to-one mapping for star-forming galaxies, but the very low sSFRs of quiescent UniverseMachine galaxies are shifted up significantly.
\label{fig:cam}}
\end{figure}

The mapping of $\mathrm{sSFR \rightarrow sSFR_{UV}}$ is not uniform because UV flux decreases with higher dust obscuration, which is strongly dependent on stellar mass \citep{Whitaker:2017}. Therefore, we map $\mathrm{sSFR \rightarrow sSFR_{UV}}$ through conditional abundance-matching (CAM) using the code \verb|conditional_abunmatch| from the \verb|halotools| package, which preserves the rank-ordering at fixed stellar mass (tolerance of $\sim$0.05~dex). We iterate this method in fuzzy redshift bins (width of $\sim$0.1) using the code \verb|fuzzy_digitize| from the \verb|halotools| package.
We match the distribution to identical photometric redshift bins of the UltraVISTA survey \citep{Muzzin:2013}.

\begin{figure}[ht!]
\includegraphics[width=0.47\textwidth]{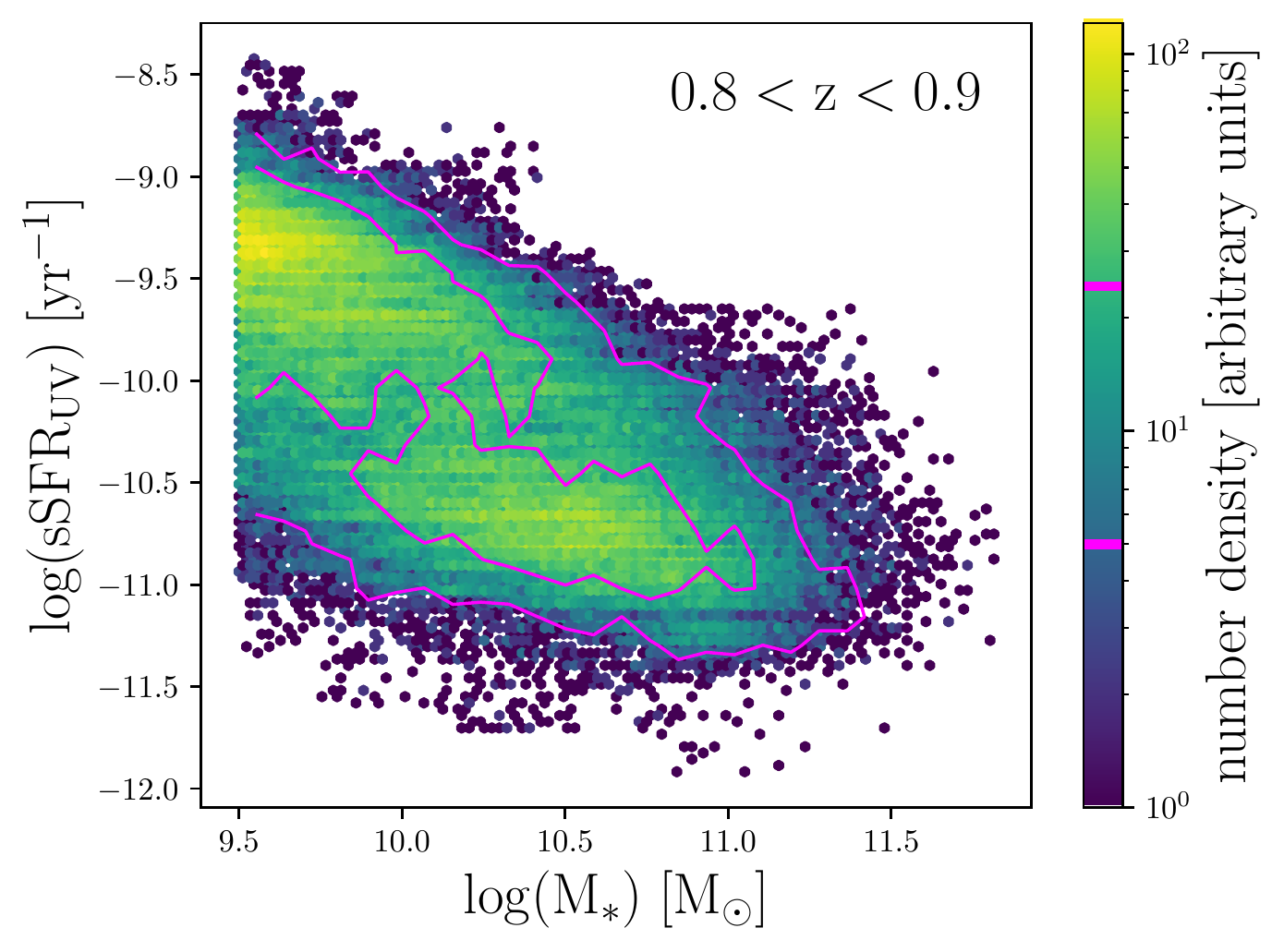}
\caption{{The 2D distribution of $M_\ast$-\ssfruv. Magenta lines represent logarithmic contours of the UltraVISTA training data, while the colored points represent logarithmic counts of UniverseMachine mock galaxies. These distributions are nearly identical by construction, via CAM. Accurately capturing this distribution is crucial for assessing mass completeness under the assumption that \ssfruv\ controls the mass-to-light ratio.}
\label{fig:m-ssfr}}
\end{figure}

\begin{figure}[ht!]
    \includegraphics[width=0.47\textwidth]{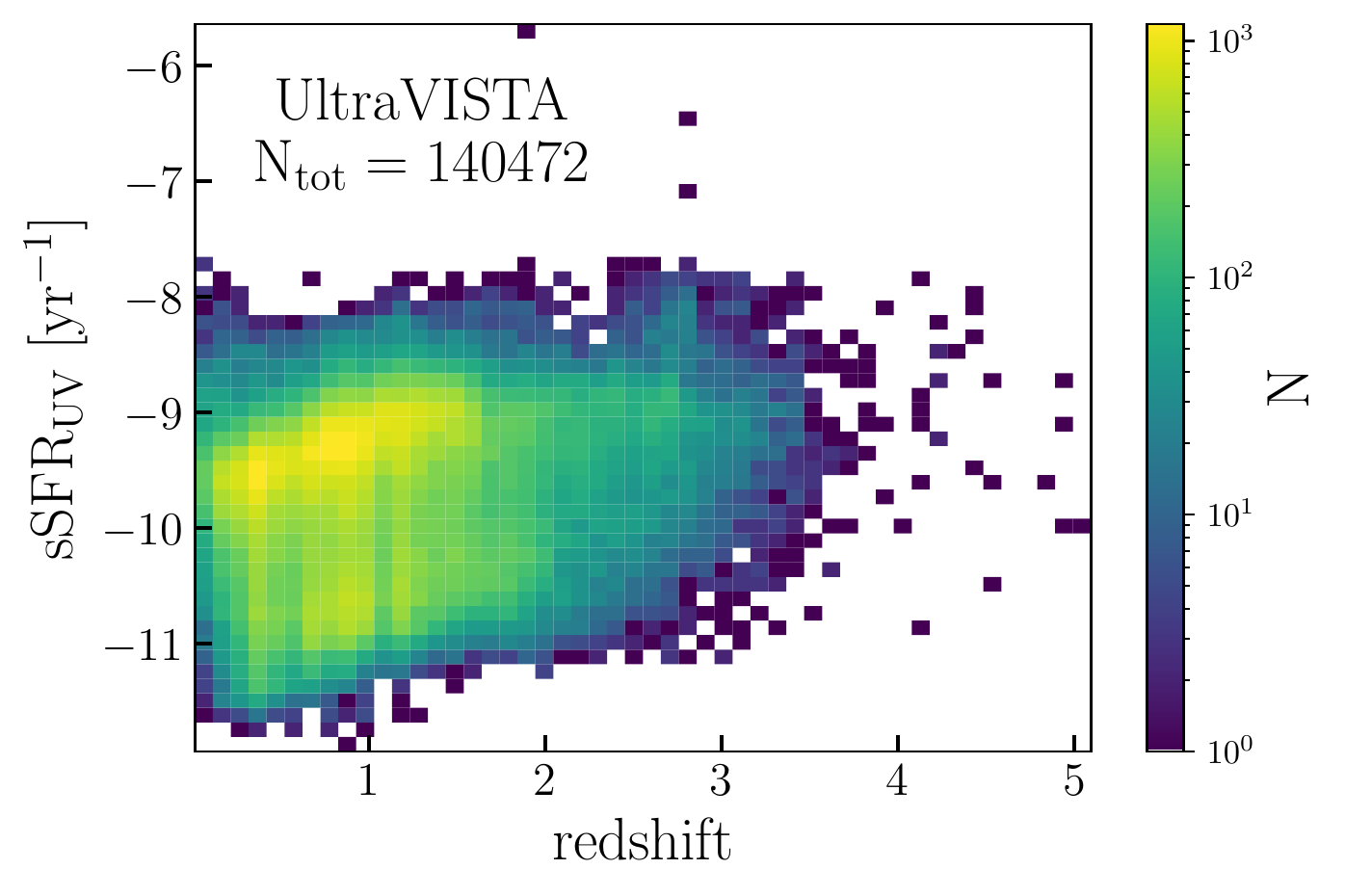}
    \caption{2D histogram of the UltraVISTA data used to train mass-to-light ratios in our random forest. There are very few regions of parameter space without any data, which makes this an ideal dataset up to $z<3$.}
    \label{fig:uvista-training-data}
\end{figure}

A visualization of the CAM mapping from the UniverseMachine $\mathrm{sSFR}$ to UltraVISTA-calibrated \ssfruv\ is shown in Figure~\ref{fig:cam}. {In the lower left panel of this figure, note that this mapping appears nearly linear for star-forming galaxies ($\rm sSFR_{tot} \gtrsim 10^{-10}$), especially at low masses where IR light is insignificant. The large gap between star-forming and quiescent galaxies ($\rm sSFR_{tot} \lesssim 10^{-10}$) is because the UniverseMachine quiescent galaxies' SFR is not fit to data past $z=0$. By construction, CAM ensures that the resulting $M_\ast$-\ssfruv\ distribution matches the data it is calibrated to (see Figure~\ref{fig:m-ssfr}), which is crucial for assessing mass completeness under the assumption that \ssfruv\ controls the mass-to-light ratio.}

\begin{figure}[ht!]
\includegraphics[width=0.47\textwidth]{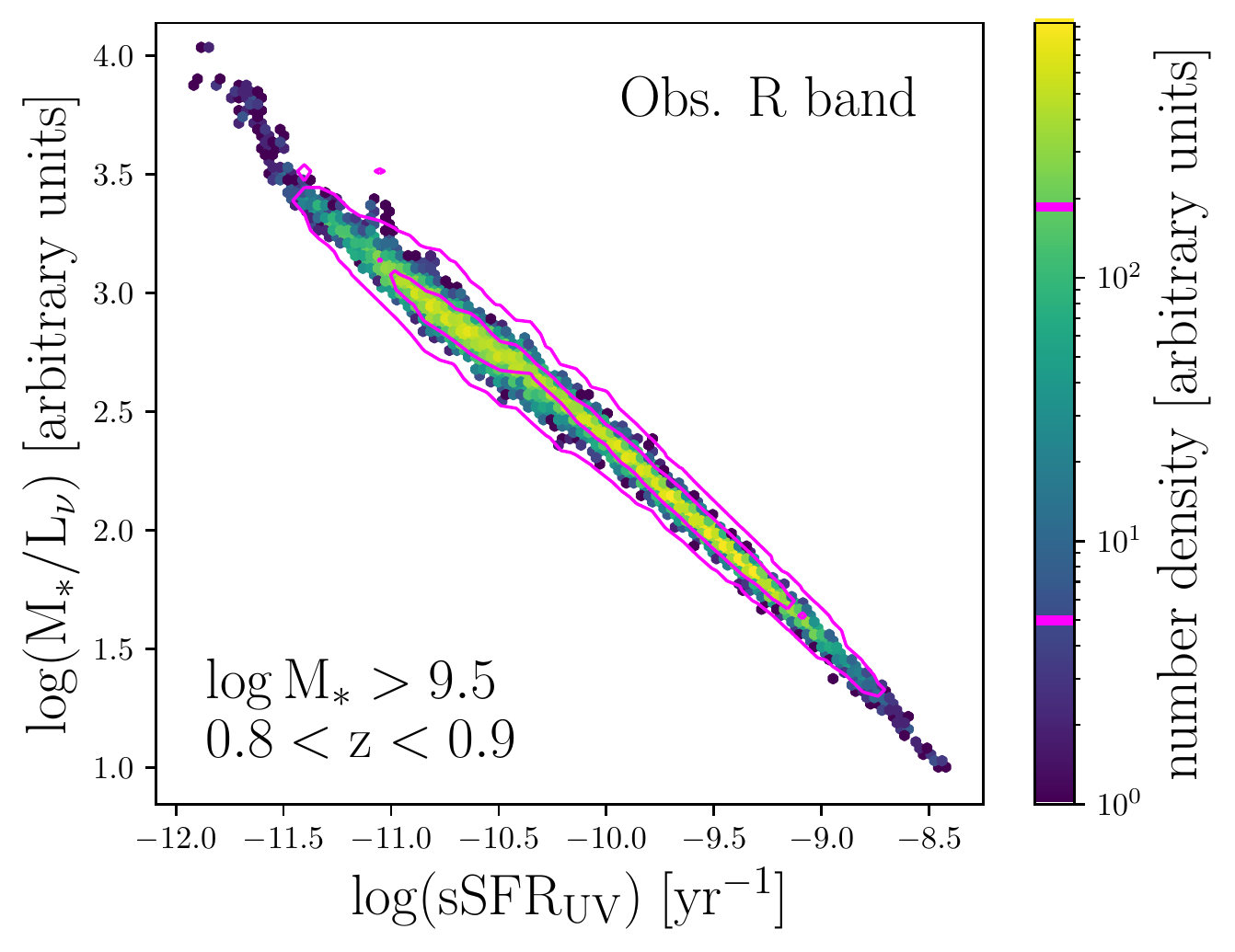}
\caption{The relation between UV specific star formation rate and mass-to-light ratio in the \textit{observed} R band at the redshift slice $0.8<z<0.9$. Magenta lines represent logarithmic contours of the UltraVISTA training data, while the colored points represent logarithmic counts of UniverseMachine mock galaxies which were fit via the Random Forest method described in Appendix~\ref{sec:climber-details}. We present the same relation in all other available photometric bands in Figure~\ref{fig:m2l-all}.
\label{fig:m2l}}
\end{figure}

\begin{figure*}[ht!]
\centering
\includegraphics[width=\textwidth]{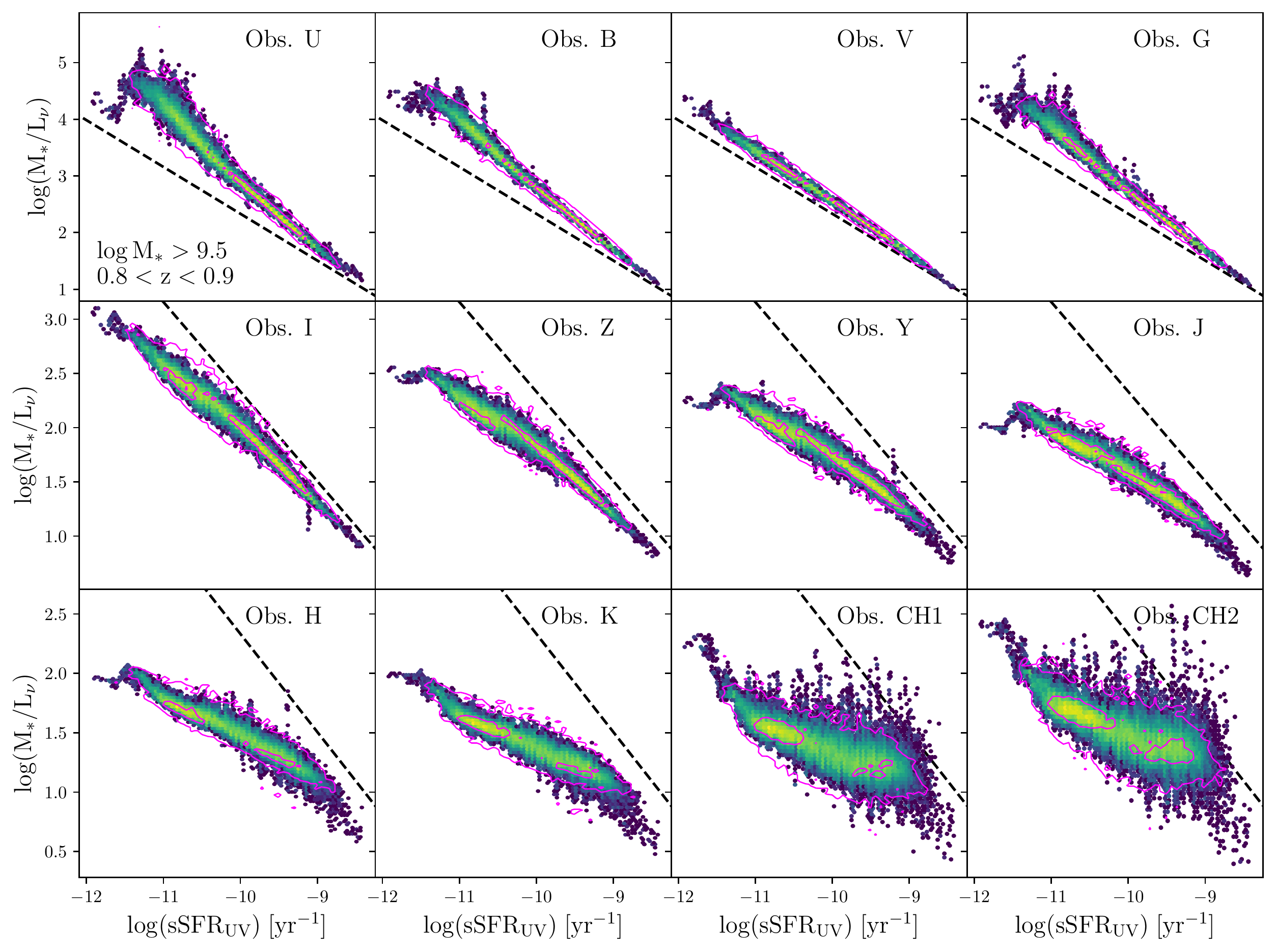}
\caption{The relation between UV specific star formation rate and mass-to-light ratio in the redshift slice $0.8<z<0.9$. Same as Figure~\ref{fig:m2l} but the y-axes represent the mass-to-light ratio in all other available photometric bands. The dashed line in each panel is the best-fit line for the R-band as a reference point.
\label{fig:m2l-all}}
\end{figure*}

We train the mapping from \ssfruv\ to mass-to-light ratio using the photometry and FAST stellar masses from UltraVISTA. We plot the feature-space of the UltraVISTA training data in Figure~\ref{fig:uvista-training-data}. These 140,{}472 training data leave very few missing regions of feature-space for $z < 3$, making it an ideal training set for our purposes. The random forest regression (the \verb|RandomForestRegressor| class from the \verb|scikit-learn| package; \citealt{Pedregosa:2012}) is then used to predict $\log(M_\ast/L)$ from the two features $\{\mathrm{\log sSFR_{UV}}, z\}$. The advantages of this approach are its simplicity, flexibility, and sufficient accuracy in predicting this relation and its intrinsic scatter (see Figure~\ref{fig:m2l} and~\ref{fig:m2l-all}). Additionally, this method automatically includes covariance between mass-to-light ratios of different photometric bands, which is important to accurately capture distributions of color as well as multivariate color-magnitude distributions. While random forests are not particularly good at extrapolation, our mock does not need to extrapolate in feature-space. Since we do not use stellar mass as a predictor of M/L (except indirectly through conditional abundance matching), it generalizes reasonably to masses slightly below the completeness limit of UltraVISTA, although predictions for galaxies far below this limit should be used with care.

\begin{figure*}[ht!]
\centering
\includegraphics[width=\textwidth]{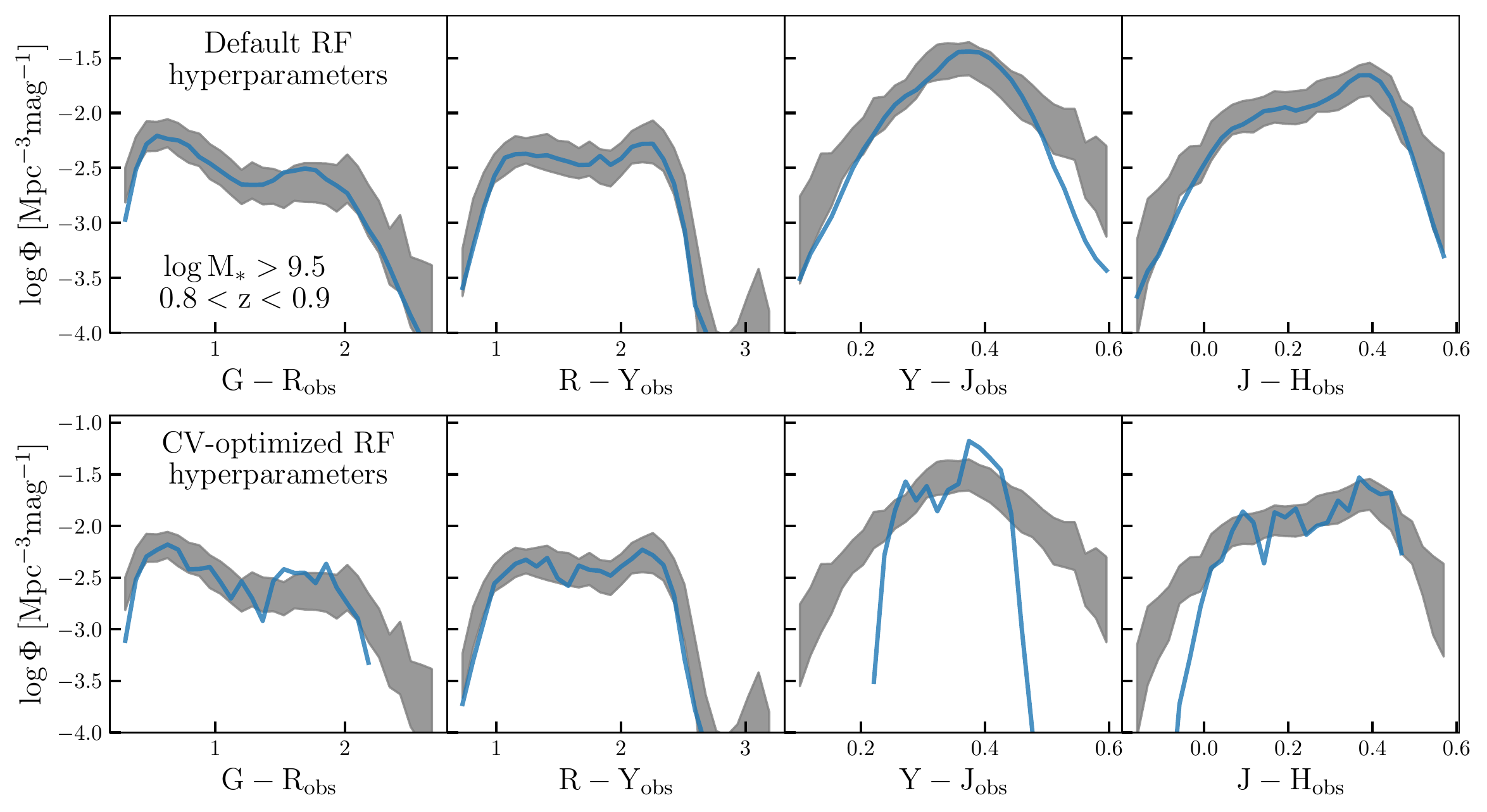}
\caption{Comparison of the color distributions between UniverseMachine mock galaxies (blue lines) and the UltraVISTA data (grey bands) that they were fit to at the redshift slice $0.8<z<0.9$. The fits in the top panel use the default scikit-learn hyperparameters, while those in the bottom panel use cross-validation-optimized hyperparameters. By construction, the optimized hyperparameters predict colors with a lower mean absolute deviation. However, we adopt the default parameters due to their significantly better performance at reproducing the distributions as a whole.
\label{fig:color-dist-vs-rf}}
\end{figure*}

To assess our mock catalog's performance at matching the UltraVISTA photometry, we compare the predicted distributions for a variety of properties, including the colors shown in Figure~\ref{fig:color-dist-vs-rf}. The mass function and luminosity functions match very well by construction, but we also want to accurately reproduce color distributions, as the selection functions in real surveys will often use color cuts in addition to magnitude cuts. We have tested a variety of choices for the random forest hyperparameters used, but found that the default scikit-learn  values performed best at reproducing a broad range of properties simultaneously. The hyperparameter values we use are therefore: \verb|n_estimators| = 10, \verb|bootstrap| = \verb|True|, \verb|max_depth| = \verb|None|, \verb|max_features| = \verb|"auto"|, \verb|min_samples_leaf| = 1, and \verb|min_samples_split| = 2.

We have applied a number of common machine learning validation methods including 5-fold cross-validation testing and learning curve analysis, using the mean absolute deviation (MAD) of colors as our loss function, and find that these hyperparameter values yield a model that is modestly overfitted (cross-validated MAD value of 0.150, versus a training score of 0.058). We calculated an alternative set of optimized hyperparameters by minimizing the mean absolute deviation of the colors (via a random hyperparameter search followed by a smaller, but exhaustive, grid search), which yielded a more converged learning curve, indicating less overfitting (cross-validated MAD value of 0.130, versus a training score of 0.126). However, while the optimized hyperparameters perform better for predictions of individual colors, the M/L values at fixed mass concentrate closely around the mean value rather than capturing the full distribution. This results in worse color distributions (as seen in the bottom panel of Figure~\ref{fig:color-dist-vs-rf}). Since we are in a regime where the feature-space distribution of the training set is nearly identical to that of the mock (by construction via conditional abundance matching), the overfitting when using the default parameters actually helps us, since it guarantees that the magnitude and color distributions in the mock match those of the training data.

\section{Additional Metrics}
\label{sec:additional-metrics}

While the two-point correlation function provides quite good HOD constraints, it is also very sensitive to cosmology; particularly the $\sigma_8$ parameter, which controls the linear bias of halos. It has been shown \citep{Slepian:2017} that the three-point correlation function would break this degeneracy, at least on linear scales. Calculating the three-point correlation function with existing public codes is too expensive to run at each iteration of an MCMC chain in our analysis. If a more efficient implementation becomes available, we would be interested in including this statistic to compare the added constraining power. This statistic will be especially important for joint analyses of cosmology and the galaxy-halo connection.

Additionally, using the two-point correlation function alone can miss out on important clustering information, as it is primarily driven by the most clustered galaxies, which is almost always satellites except at the highest masses. One can explicitly measure a signal from central galaxies singling them out via isolation criteria or group catalog reconstruction. One metric in particular that is capable of measuring assembly bias signals is the number of satellites vs. central stellar mass, split into quenched and star-forming populations (shown in private communication with Rodr\'{i}guez-Puebla). It is unclear how accurately group catalogs can be reconstructed from surveys like PFS and MOONS, which are incomplete due to fiber collisions, and require the use of photometric redshifts to fill in the gaps. This may prove to be a strong argument for extensions to these surveys prioritizing increasing the {targeting} completeness over the area, contrary to the cosmic variance tests herein.

Alternatively, it is possible to supplement the two-point correlation function with other statistics such as counts-in-cylinders (CIC). By tuning the radius of the cylinder and inner annulus, CIC can effectively separate the signal of centrals from satellites, without requiring full group catalog reconstruction. \citet{Wang:2019} demonstrate that CIC + $w_{\rm p}(r_{\rm p})$ may be sufficient to constrain assembly bias as well. However, further testing is required for samples affected by fiber collisions like PFS and MOONS.

\end{document}